\documentclass[a4paper,11pt]{amsart}

\usepackage{amsmath}
\usepackage{amssymb}
\usepackage{euscript}
\usepackage{epsfig}

\usepackage{booktabs}

\newtheorem{Thm}{Theorem}
\newtheorem{Lem}[Thm]{Lemma}
\newtheorem{Prop}[Thm]{Proposition}
\newenvironment{Proof}{\begin{proof}}{\end{proof}}

\newcommand{\bangle}[2]{{{#1}\atopwithdelims\langle\rangle{#2}}}

\newcommand{\myF}{\ensuremath{\EuScript F}}

\newcommand{\myA}{\ensuremath{\EuScript A}}

\newcommand{\myL}{\ensuremath{\EuScript L}}
\newcommand{\myB}{\ensuremath{\EuScript B}}
\newcommand{\myP}{\ensuremath{\EuScript P}}
\newcommand{\myR}{\ensuremath{\EuScript R}}

\newcommand{\F}{\ensuremath{\mathbb F}}
\newcommand{\poly}{\ensuremath{\mathrm{poly}}}
\newcommand{\sgn}{\ensuremath{\operatorname{sgn}}}

\long\def\comment#1{}

\usepackage{tikz}
\usetikzlibrary{decorations.markings}
\usetikzlibrary{arrows}

\def\sixpacksmall{%
\begin{scope}[font=\small\it\color{blue}]
\node[vertex,label=above:a] (A) at (0,0) {};
\node[vertex,label=above:b] (B) [right of = A] {};
\node[vertex,label=above:c] (C) [right of = B] {};
\node[vertex,label=below:d] (D) [below of = A] {};
\node[vertex,label=below:e\vphantom{f}] (E) [right of = D] {};
\node[vertex,label=below:f] (F) [right of = E] {};
\draw (A)--(B)--(C)--(F)--(E)--(D)--(A);
\draw (D)--(B)--(F); \draw (B)--(E);
\end{scope}
}

\def\exgraph{\sixpacksmall}

\newcommand{\edge}[1]{\ensuremath\text{\it #1}}

\begin{document}


\title[]{Narrow sieves for\\parameterized paths and
  packings}
\author[]{\vspace*{-0.1in}Andreas Bj\"{o}rklund$^1$}
\address{%
$^1$Lund University,
Department of Computer Science,
P.O.Box 118, SE-22100 Lund, Sweden}
\email{andreas.bjorklund@yahoo.se}
\author[]{Thore Husfeldt$^{2}$}
\address{%
$^2$IT University of Copenhagen,
2300 Copenhagen S, Denmark and Lund University,
Department of Computer Science,
P.O.Box 118, SE-22100 Lund, Sweden. Partially supported by Swedish
Research Council grant 2007-6595.}
\email{thore@itu.dk}
\author[]{Petteri Kaski$^3$}
\address{%
$^3$Helsinki Institute for Information Technology HIIT,
Aalto University, 
Department of Information and Computer Science
PO Box 15400, FI-00076 Aalto, Finland}
\email{petteri.kaski@tkk.fi}
\author[]{Mikko Koivisto$^4$}
\address{%
$^4$Helsinki Institute for Information Technology HIIT,
Department of Computer Science, University of Helsinki,
P.O.Box 68, FI-00014 University of Helsinki, Finland. Partially
supported by Academy of Finland grant 125637.}
\email{mikko.koivisto@cs.helsinki.fi}

\begin{abstract}
  We present randomized algorithms for some well-studied, hard
  combinatorial problems: the $k$-path problem, the $p$-packing of
  $q$-sets problem, and the $q$-dimensional $p$-matching problem. Our
  algorithms solve these problems with high probability in time
  exponential only in the parameter ($k$, $p$, $q$) and using
  polynomial space; the constant bases of the exponentials are
  significantly smaller than in previous works. For example, for the
  $k$-path problem the improvement is from 2 to 1.66. 
  We also show how to detect if a $d$-regular graph admits an edge
  coloring with $d$ colors in time within a polynomial factor of
  $O(2^{(d-1)n/2})$.

  Our techniques build upon and generalize some recently published
  ideas by I. Koutis (ICALP 2009), R. Williams (IPL 2009), and
  A. Bj\"ork\-lund (STACS 2010, FOCS 2010).
\end{abstract}


\thispagestyle{empty}

\newlength{\parindentsave}
\setlength{\parindentsave}{\parindent}
\newlength{\parskipsave}
\setlength{\parskipsave}{\parskip}
\setlength{\parindent}{0pt}
\setlength{\parskip}{0pt}

\maketitle


\setlength{\parindent}{\parindentsave}
\setlength{\parskip}{\parskipsave}

\renewcommand{\thepage}{\arabic{page}}


\section{Introduction}

Combinatorial problems such as finding a long simple path in a graph
or disjointly packing many members of a set of subsets are
well-studied and hard. In fact, under standard complexity-theoretic
assumptions, algorithms for these problems must either be inexact, or
require running times of super-polynomial or maybe even exponential
time.

It has been observed that the complexity of the problems studied in
the present paper depends exponentially on the output size $k$ instead
of the input size $n$, i.e., they admit running times of the form
$\exp(\poly(k))\cdot\poly(n)$ rather than, say, $\exp(O(n))$. A number of
papers have improved the exponential dependencies dramatically over
the past decade, arriving at exponential factors of size $2^k$. We
further improve this dependency, reducing the exponent base below the
constant 2, sometimes significantly.

To express our results in the terminology of parameterized
computational complexity theory, we improve the running time of
several canonical fixed parameter tractable problems. In particular,
we claim the ephemeral lead in the highly competitive ``FPT races''
for path finding, uniform set packing, and multidimensional matching.

\medskip
Our techniques also allow us to report progress on an unparameterized
problem: we show a nontrivial upper bound on the complexity of edge
coloring for regular graphs.

\bigskip We adopt the notational convention from parameterized and
exponential time algorithms, letting $O^*(f(k))$ denote
$f(k)n^{O(1)}$, where typically $n$ is some aspect of the input size
such as number of vertices, and $k$ is a parameter such as path
length. Our parameters are all polynomially bounded by $n$, so $O^*$
also hides factors that are polynomial in the parameter size. We
present our results in terms of decision problems, they can be turned
into optimization or search problems by self-reductions in the obvious
way.

All our algorithms are randomized. The error is one-sided in the sense
that they never report a false positive. The error probability is
constant and can be made exponentially small by a polynomial number of
repetitions, which would again be hidden in the $O^*$ notation.

\subsection{Finding a path} Given an undirected graph $G$ on $n$ vertices, the
\emph{$k$-path problem} asks whether $G$ contains a
simple path on $k$ vertices. 
\begin{Thm}
  \label{thm: path}
  The undirected $k$-path problem can be solved in time $O^*(1.66^k)$ by
  a randomized algorithm with constant, one-sided error.
\end{Thm}

The proof is in \S\ref{sec: k-path}. For $k=n$, the result matches the
running time of the recent algorithm for Hamiltonian path of
Bj\"orklund \cite{Bjork2010b}.

\subsubsection*{Previous work.} 
Naively, the $k$-path problem can be solved in time $O^*(n^k)$, but
Monien \cite{Monien1985} and Bodlaender \cite{Bodlaender1993} showed
that the problem can be solved in time $O^*(f(k))$, leading
Papadimitriou and Yannakakis \cite{PapaYanna} to conjecture that the
problem was polynomial-time solvable for $k=O(\log n)$. This was
confirmed in a strong sense by Alon, Yuster, and Zwick, with a
beautiful $O^*(c^k)$ algorithm. A number of paper have since reduced
the base $c$ of the exponent using different techniques, see Table~1. (In the following tables, we mark
randomized algorithms with `r'.)

\begin{table}[htb]
{\small
\begin{tabular}{rclrcl}\toprule
\multicolumn{3}{l}{Table 1. $k$-path in time $O^*(f(k))$}\\\midrule
 $k!$ & &Monien \cite{Monien1985} &
   $12.6^k$ & & Chen \emph{et al.} \cite{Chen2007} \\
 $k!2^k$ & &Bodlaender \cite{Bodlaender1993}  &
   $4^k$ & r & Chen \emph{et al.} \cite{Chen2007} \\
  $5.44^k$ & r &Alon \emph{et al.} \cite{AYZ1995} \hspace*{2cm}&
   $2.83^k$ & r & Koutis (2008) \cite{Koutis2008}\\
   $c^k$ & & $c>8000$, Alon \emph{et al.}  \cite{AYZ1995}  &
   $2^k$ & r &  Williams \cite{Williams2009} \\
   $16^k$ & & Kneis \emph{et al.} \cite{Kneis2006} &
   $1.66^k$ & r & \emph{this paper}\\\bottomrule
\end{tabular}
}
\end{table}

Our result is yet another improvement of the exponent base, notable
perhaps mostly because it breaks the psychological barrier of
$c=2$. We fully expect this development to continue. On the other
hand, computational complexity informs us that an even more ambitious
goal may be quixotic: An algorithm for $k$-path with running time
$\exp(o(k))$ would solve the Hamiltonian path problem in time
$\exp(o(n))$, which is known to contradict the exponential time
hypothesis \cite{IPZ}.

\subsection{Packing disjoint triples}
\label{sec: intro triple packing}
Let $\myF$ be a family of subsets of an $n$-element ground set. A
subset $\myA\subseteq\myF$ is a $p$-packing if $|\myA|=p$ and the sets
in $\myA$ are pairwise disjoint. Given input set $\myF$ of size-3
subsets, the \emph{$p$-packing of $3$-sets} problem asks whether
$\myF$ contains a $p$-packing.  This problem includes a number of
well-studied problems in which the ground set consists of the
vertices of an input graph $G=(V,E)$.  In the \emph{vertex-disjoint
  triangle $p$-packing problem}, the set $\myF$ consists of the
subsets of $V$ that form a triangle $K_3$. In the \emph{edge-disjoint
  triangle $p$-packing problem}, the set $\myF$ consists of the
subsets of $E$ that form the edges of a triangle. In the
\emph{vertex-disjoint $P_3$ $p$-packing problem}, the set $\myF$
consists of the vertex subsets $\{u,v,w\}\subseteq V$ for which
$uv,vw\in E$.

\begin{Thm}\label{thm: triple packing}
  The $p$-packing of $3$-sets problem can be solved  by a randomized
  algorithm in time
  $O^*(1.493^{3p})$ with constant one-sided error.
\end{Thm}

The proof is in \S\ref{sec: packing}. For $p=n$, the result matches
the running time of the recent algorithm for exact cover by 3-sets of
Bj\"orklund \cite{Bjork2010a}.

\subsubsection*{Previous work} The naive algorithm for $p$-packing
considers all $\binom{|\myF|}{p}$ ways of selecting $p$ sets from
$\myF$.  Results of the form
$O^*(f(p))$ go back to Downey and Fellows \cite{DowneyFellows1999},
and the dependency on $p$ has been improved dramatically in a series
of papers, see Table~2. Remarkably, the best previous running time is given by
Koutis's algorithm for the more general problem of $p$-packing sets of
size $q$, specialized to the case $q=3$. (We return to the performance
of our own algorithm in the case $q>3$ in \S\ref{sec: intro q-sets}.)

\begin{table}[htb]
{\small
\begin{tabular}{lcl}\toprule
\multicolumn{3}{l}{Table 2: $p$-packings of $3$-sets in time $O^*(f(p))$}\\\midrule
  $2^{O(p)}(3p)!$ & & Downey and Fellows \cite{DowneyFellows1999}\\
 $(5.7p)^p$ & r & Jia, Zhang, and Chen \cite{Jia2004} \\
$2^{O(p)}$ & & Koutis (2005) \cite{Koutis2005} \\
 $(12.7 c)^{3p}$ & & $c>10.4$, Fellows, Heggernes, \emph{et al.} \cite{FellowsPinar}
 \\
$10.88^{3p}$ & r & Koutis (2005) \cite{Koutis2005}\\
 $4.68^{3p}$ & & Kneis et al. \cite{Kneis2006} \\
 $4^{3p+o(p)}$ & & Chen et al. \cite{Chen2007}\\
 $2.52^{3p}$ & r &  Chen et al. \cite{Chen2007}\\
  $2^{2p\log p+ 1.869p}$ & &vertex-disjoint triangles $K_3$, Fellows, Heggernes,
  \emph{et al.} \cite{FellowsPinar}$^\dag$
  \\
  $22.628^{p\log p + p}$& &edge-disjoint triangles $K_3$,
  Mathieson \emph{et al.} \cite{Mathieson2004}\\
  $4.61^{3p}$ & & Liu et at. \cite{Liu}\\
  $3.523^{3p}$ & & Wang and Feng \cite{WangFeng} \\
  $3.404^{3p}$ & & vertex-disjoint paths $P_3$, Prieto and
  Sloper \cite{PrietoSloper2006}\\
  $2.604^{3p}$ & & vertex-disjoint paths $P_3$, Wang et al. \cite{Wang2008}\\
  $2.482^{3p}$ & &  vertex-disjoint paths $P_3$, Fernau and Raible \cite{FernauRaible}\\
  $2^{3p}$ & r&   Koutis \cite{Koutis2008}\\
  $1.493^{3p}$ & r&   \emph{this paper}\\\bottomrule
\multicolumn{3}{l}{$^{\dag\vphantom{^f}}$ The precise time bound can be seen to be $O(2^{2p\log p+1.869p} n^3)$.}
\end{tabular}
}
\end{table}

It is known that packing vertex-disjoint copies of $H$ into $G$ is
NP-complete as soon as $H$ is connected and has more than 2
vertices\cite{HellKirkpatrick}. Fellows \emph{et al.}
\cite{FellowsPinar} raised the question of how hard the parameterized
problem is, and we observe here that this can be answered under the
exponential time hypothesis \cite{IPZ}, which is equivalent to the
parameterized complexity hypothesis $\text{FPT}\neq \text{M}[1]$.

\begin{Prop}\label{prop: triangle packing eth hard}
  There is no algorithm for vertex-disjoint triangle $p$-packing in
  time $\exp(o(p))$ unless the satisfiability of 3-CNF formulas in $n$
  variables can be decided in time $\exp(o(n))$.
\end{Prop}

The proof of the proposition is routine and we omit a detailed
presentation. Briefly, a 3-CNF instance to satisfiability is
transformed to a 3-dimensional matching instance; the standard
reduction results in a graph of size $O(n+m)$, and the sparsification
lemma of Impagliazzo, Paturi, and Zane \cite{IPZ} is used to remove
the dependency on $m$. A subexponential time (in $p\leq n$) algorithm
for this problem would solve the original instance in time
$\exp(o(p))=\exp(o(n))$.  

\subsection{Uniform set packing}
\label{sec: intro q-sets}

As before, let $\myF$ be a set of subsets of an $n$-element ground
set. A subset $\myA\subseteq\myF$ is a $p$-{\em packing} if $|\myA|=p$
and the sets in $\myA$ are pairwise disjoint.  Given as input a
family $\myF$ of size-$q$ subsets, the \emph{$p$-packing of $q$-sets}
problem asks us to determine whether $\myF$ contains a $p$-packing.

This is a generalization of the triple $p$-packing problem described
in \S\ref{sec: intro triple packing}, and the algorithm we advertised
in that section is merely a specialization of a more general
result. 

\begin{Thm}\label{thm: q-set packing}
  The $p$-packing of $q$-sets problem can be solved by a randomized algorithm in time
  $O^*(f(p,q))$ with constant, one-sided
  error, where
\[ 
f(p,q)= \left\{ \frac{0.108157\cdot
    2^q(1-1.64074/q)^{1.64076-q}q^{0.679625}}{
    (q-1)^{0.679623}}\right\}^p \,.\]
\end{Thm}

Potentially, $\myF$ can have size $\binom{n}{q}$, so reading in the
input alone can take super-polynomial time in $n$. Thus we adopt the
convention that $|\myF|$ is polynomial in $n$. Thus, the $O^*$
notation suppresses polynomial factors in $n$ (and hence in $p,q\leq
n$) and also in $|\myF|$; a more careful (and even less readable)
bound on the running time is given in \S\ref{sec: p-packings
  of q-sets algorithm}.

Still, the above expression is difficult to parse. The previous best
bound is Koutis's \cite{Koutis2008} much cleaner $O^*(2^{qp})$, and
our bound is not $O^*((2-\epsilon)^{qp})$ for any $\epsilon$. Instead,
our algorithm behaves well on small $q$; for comparison, we can
express bounds on $f$ of the form $O^*(c^{qp})$ for small $q\geq 3$.

\medskip
{\small\begin{tabular}{ccccccccc}
$q$& $3$ &   $4$ & $5$ & $6$ & 7 & 8 \\\midrule
$f(p,q)$ & $1.4953^{3p}$ & $1.6413^{4p}$ & $1.7205^{5p}$ & $1.7707^{6p}$  &
$1.8055^{7p}$ & $1.8311^{8p}$
\end{tabular}

\medskip
\begin{tabular}{ccccccc}
$q$& $10$ &   $20$ & $50$ & $100$ & $500$ \\\midrule
$f(p,q)$ & $1.8663^{10p}$ & $1.9345^{20p}$ & $1.9741^{50p}$ &
$1.9871^{100p}$ & $1.9975^{500p}$
\end{tabular}}

\bigskip
The proof is in \S\ref{sec: packing}. For $p=n$, the result matches the running time of the recent algorithm for
exact cover by $q$-sets of Bj\"orklund \cite{Bjork2010a}.

\subsubsection*{Previous work}

Most of the work on $p$-packing of $q$-sets has been done for the
special case $q=3$, described in \S\ref{sec: intro triple
  packing}. For general $q$, the first algorithm with running time of
the form $O^*(f(p,q))$ is due to Jia \emph{et al.} \cite{Jia2004}; the
subsequent improvements are shown in Table~4.

\begin{table}[htb]
\small
\begin{tabular}{lcl}\toprule
\multicolumn{3}{l}{Table 4: $p$-packing of $q$-sets in time
  $O^*(f(p,q))$}\\\midrule
$\exp(O(pq))$& & Fellows, Knauer, \emph{et al.} \cite{FellowsKnauer}\\
 $5.44^{qp}$ & r & Koutis \cite{Koutis2005}\\
 $2^{qp}$ & r & Koutis \cite{Koutis2008}\\\bottomrule
\end{tabular}
\end{table}

For the special graph packing case where $\myF$ consists of isomorphic
copies of a fixed graph $H$ of size $q$, an earlier result
established $O(2^{pq\log p+2pq\log q}n^q)$ \cite{FellowsPinar}.
The only specific packing problem we are aware of that our general
algorithm does not seem to improve is the problem of packing
vertex-disjoint stars into a given graph. A star $K_{1,q-1}$ consists
of a center vertex connected to $q-1$ other vertices. Prieto and
Sloper \cite{PrietoSloper2006} exhibit a kernel of polynomial size
$s(p,q)=p(q^3+pq^2+pq+1)$ for this problem. They do not express
running times in terms of $q$, but their ``brute force'' algorithm can
be seen to run in time within a polynomial factor
of \(\binom{s(p,q)}{p} = \exp(O(p\log pq)).\)

Significant improvements for $p$-packing of $q$-sets, such as a
general $\exp(q\cdot o(p))$ algorithm, are ruled out by
Proposition~\ref{prop: triangle packing eth hard}. For the nonuniform
$p$-packing problem, when there is no bound on the size of the packed
sets, we are unlikely to find an algorithm with running time
$O^*(f(p))$ for any function $f$; the main evidence is provided in
terms of parameterized complexity, where the more specific problem of
finding an independent set of size $p$ (equivalently, packing $p$
subsets of is the family $\myF$ of closed vertex neighborhoods in a
given graph) is $W[1]$-hard \cite{DowneyFellows1999}.

\subsection{Multidimensional matching}

Let $U_1,U_2,\ldots,U_q$ be pairwise disjoint sets, each of size
$r$. Let $\myF$ be a set of subsets of $U_1\cup U_2\cup\cdots\cup U_q$
such that each $A\in\myF$ satisfies $|A\cap U_j|=1$ for each
$j=1,2,\ldots,q$.  Given $\myF$ as input, the {\em $q$-dimensional
  $p$-packing problem} asks whether $\myF$ contains a
$p$-packing. One often views this problem as $q$-dimensional
\emph{$p$-matching}, in which case $\myF$ is thought of as the edge
set of a $q$-uniform $q$-partite hypergraph over $U_1\times U_2\times
\cdots\times U_q$. 

Again, $\myF$ itself can have size up to $r^q$, so reading the
input alone would take time super-polynomial in the size $n=qr$ of the
universe already. Thus, we adopt the convention that $|\myF|$ is
polynomial in $n$. In particular, the $O^*$ notation hides factors
polynomial in both $n$ (and hence $p,q\leq n$) and $|\myF|$.

\begin{Thm}\label{thm: matching}
  The $q$-dimensional $p$-packing problem can be solved in time
  $O^*(2^{(q-2)p})$ by a randomized algorithm with constant, one-sided error.
\end{Thm}
The proof is in \S\ref{sec: matching}.
For $p=r$, the result matches the running time of the recent algorithm for
$q$-dimensional perfect matching of Bj\"orklund \cite{Bjork2010a}.

\subsubsection*{Previous work.} The first parameterized
multi-dimensional matching algorithm appears to be Downey, Fellows,
and Koblitz's \cite{DFK} application of the color-coding
technique. Some of the ensuing improvements apply only to the
3-dimensional case. See Table 3.

\begin{table}[htb]\small
\begin{tabular}{lcl}\toprule
\multicolumn{3}{l}{Table 3: $q$-dimensional $p$-packing in time
  $O^*(f(p,q))$}\\\midrule
$(qp)!(qp)^{3qp+1}$ & & Downey, Fellows, Koblitz \cite{DFK}\\
$\exp(O(pq))$ & & Fellows, Knauer, et al. \cite{FellowsKnauer}\\
$\exp(O(pq))$ & & Koutis (2005) \cite{Koutis2005}\\
$(4\mathrm e)^{pq}$ & r & Koutis (2005) \cite{Koutis2005}\\
$2.80^{3p}$ & &$q=3$,  Chen et al. \cite{Chen2007}\\
$2.77^{3p}$ & &  $q=3$, Liu \emph{et al.} \cite{Liu}\\
$2.52^{3p}$ & r &$q=3$,  Chen et al. \cite{Chen2007}\\
$2^{qp}$ & r & Koutis (2008) \cite{Koutis2008}\\
  $2^{(q-1)p}$ & r & Koutis and Williams \cite{KoutisWilliams2009}\\
  $2^{(q-2)p}$ & r & \emph{this paper}\\\bottomrule
\end{tabular}
\end{table}

\subsection{Edge colouring}
Finally, we turn to an unparameterized problem.

Let $G$ be an undirected loopless $n$-vertex $d$-regular graph without
parallel edges.  The {\em edge-coloring problem} asks whether the
edges of $G$ can be colored so that no two edges that share an
endvertex have the same color. It is well known that the number of
colors required is either $d$ or $d+1$.

\begin{Thm}\label{thm: edge coloring}
  The $d$-edge coloring problem for $d$-regular graphs can
  be solved in time $O^*(2^{(d-1)n/2})$ and polynomial space by a
  randomized algorithm with constant, one-sided error.
\end{Thm}

The proof is in \S\ref{sec: coloring}.

\subsubsection*{Previous work} The naive algorithm for edge coloring
tests all $d^m$ assignments a $d$ colors to $m$ edges. To the best of our
knowledge, the only other known exact algorithm is to look for a
\emph{vertex} $d$-coloring of the line graph $L(G)$ of $G$. The line
graph $L(G)$ has $m$ vertices, so the algorithm of Bj\"orklund,
Husfeldt, and Koivisto \cite{BHK} solves the problem in time (and space)
$O^*(2^m)$, which is $O^*(2^{dn/2})$ for $d$-regular graphs. For
$3$-coloring, the current best bound is $O(1.344^n)$
\cite{Kowalik2009}; very recently, exponential
space enumeration algorithms for 3, 4, and 5-edge colouring were
announced \cite{GKC}, with running time $O(1.201^n)$, $O(1.8172^n)$,
and $O(3.6626^n)$, respectively.

It is known that for each $d\geq 3$ it is an NP-complete problem to
decide whether $d$ colors suffice \cite{HolyerColor,LevenGalil}. However, the exponential time
complexity of edge coloring remains wide open \cite{Kowalik2008}. Curiously, we do not know how to apply these ideas
to \emph{vertex} coloring; a polynomial space algorithm with running
time $O^*(2^n)$ has not yet been found.

\subsection{Methods}

Our methods follow the idea introduced by Koutis \cite{Koutis2008} of
expressing a parameterized problem in an algebraic framework by
associating multilinear monomials with the combinatorial structures we
are looking for, ultimately arriving at a polynomial identity testing
problem. Various ideas are used for sieving through these monomials by
canceling unwanted contributions.

Some of our results are the parameterized analogues of recent work by
the first author \cite{Bjork2010a, Bjork2010b}, all using a
determinant summation idea that essentially goes back to Tutte. To
make these ideas work in the parameterized setting is not
straightforward. In particular, while the Hamiltonian path algorithm
\cite{Bjork2010b} uses determinants to cancel the contribution of
unwanted labeled cycle covers, our $k$-path algorithm from
Theorem~\ref{thm: path} uses a
combinatorial argument to pair unwanted labeled walks of $k$
vertices. With $k=n$ we recover a $O^*(1.66^n)$ Hamiltonian path
algorithm, but using a different (and arguably more natural) approach.
On the other hand, the parameterized packing and matching results of
Theorems~\ref{thm: triple packing}, \ref{thm: q-set packing}, and
\ref{thm: matching}, and also the edge colouring result in
Theorem~\ref{thm: edge coloring} all use determinants.

Our algorithm for $k$-path seems to be subtle in the sense that we see
no way of extending it to other natural combinatorial structures, even
directed paths. In contrast, the ideas of Koutis \cite{Koutis2008} and
Williams \cite{Williams2009} work directed graphs, and for detecting
$k$-vertex trees and $k$-leaf spanning trees \cite{KoutisWilliams2009}
in time $O^*(2^k)$.

\medskip

It seems to be difficult to achieve our results using previous
techniques.  In particular, the limitations of the group algebra
framework \cite{Koutis2008, Williams2009} were studied by Koutis and
Williams \cite{KoutisWilliams2009}; they show that multilinear
polynomials of degree $k$ cannot be detected in time faster than
$O^*(2^k)$ in their model.  Koutis \cite{Koutis2008} has argued that
the color coding method  \cite{AYZ1995} and the randomized
divide-and-conquer approach  \cite{Chen2007}) also cannot achieve
running times whose exponent base is better than $O^*(2^k)$.

\section{A Projection Sieve for $k$-Paths}
\label{sec: k-path}

This section establishes Theorem~\ref{thm: path}.

\subsection{Overview}
\label{sec: path overview}
In this section, we develop an inclusion--exclusion sieve 
over multivariate polynomials for the $k$-path problem.

The input graph is randomly partitioned into two sets $V_1,V_2$ of
roughly equal size. Central to our analysis is the family of
\emph{labeled walks}, defined relative to such a random partition as
follows: Each occurrence on the walk of a vertex in $G[V_1]$ and of an
edge in $G[V_2]$ receives a unique label. We call this a
\emph{bijective} labeling. The labels need not correspond to the
order in which the objects are visited, and the same object can incur
more than one label. For example, with $V_1=\{a,b,c\}$, $V_2=\{d,e,f\}$,

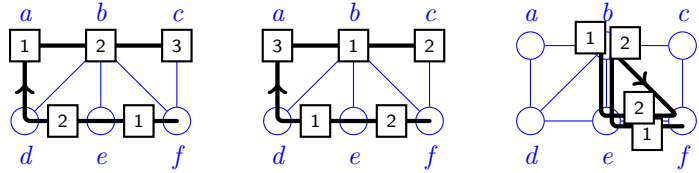
\def\arrowposition{.5}
\tikzstyle{vertex}=[draw,circle,minimum size=10pt]
\tikzstyle{lbl}=[
  font=\sf\tiny,rectangle,draw,thick,fill=white,sharp corners]
\tikzstyle{walk}=[decoration={markings,mark=at position \arrowposition with{\arrow{to};}},ultra thick, rounded corners=2pt, postaction={decorate},line cap=round]

\[
\vcenter{\hbox{
\begin{tikzpicture}
  \begin{scope}[draw=blue]\exgraph\end{scope}
 \draw[walk]
 (F.center)--node[lbl]{1} 
 (E.center)--node[lbl]{2}
 (D.center)--
 (A.center)--node[at start, lbl]{1}
 (B.center)--node[at start, lbl]{2} node[at end, lbl]{3}
 (C.center);
\end{tikzpicture}}}
\qquad
\vcenter{\hbox{
\begin{tikzpicture}
 \begin{scope}[draw=blue]\exgraph\end{scope}
 \draw[walk]
 (F.center)--node[lbl]{2} 
 (E.center)--node[lbl]{1}
 (D.center)--
 (A.center)--node[at start, lbl]{3}
 (B.center)--node[at start, lbl]{1} node[at end,lbl]{2}
 (C.center);
\end{tikzpicture}}}
\qquad
\vcenter{\hbox{
\begin{tikzpicture}
 \begin{scope}[draw=blue]\exgraph\end{scope}
 \draw[walk]
 ([shift={(0pt,-2pt)}]F.center)-- node[lbl,yshift=-3pt]{1}
([shift={(2pt,-2pt)}]E.center)-- node[lbl, at end, shift={(5pt,3pt)}]{2}
 ([shift={(2pt,-2pt)}]B.center)--
 ([shift={(-2pt,2pt)}]F.center)-- node[lbl,yshift=3pt]{2}
 ([shift={(-2pt,2pt)}]E.center)-- node[lbl, at end,shift={(-4pt,1pt)}]{1}
 ([shift={(-2pt,2pt)}]B.center);
\end{tikzpicture}}}
\]

Note the asymmetry between what is labeled in $G[V_1]$ and $G[V_2]$;
indeed, with good probability a path of length $k$ has roughly $k/2$
vertex labels in $G[V_1]$ but only $k/4$ edge labels in $G[V_2]$. Very
roughly speaking, the running time of our algorithm is around
$2^{k/2+k/4}$ for that reason.

With each such labeled walk we associate a monomial consisting of
variables $x_e$ for every edge $e$ on the walk, variables $y_{v,i}$
for every vertex $v\in G[V_1]$ labeled $i$, and variables $z_{e,i}$
for every edge $e\in G[V_2]$ labeled $i$.  For example, the monomial
associated with leftmost labeled walk above is
\[x_{\edge{ab}}\cdot x_{\edge{ad}}\cdot x_{\edge{bc}}\cdot
x_{\edge{de}}\cdot x_{\edge{ef}}\cdot y_{{a},1} \cdot
y_{{b},2}\cdot y_{ c,3} \cdot z_{{de},2}\cdot
z_{\edge{ef},1}\,;\]the monomial associated with the middle labeled
walk contains the same $x$s, but the remaining factors are \(
y_{a,3} \cdot y_{b,1}\cdot y_{c,2} \cdot
z_{\edge{de},1}\cdot z_{\edge{ef},2}\).

If the labeled walk is a path (i.e., it has no repeated vertices),
then it can be uniquely recovered, including the labels, from its
associated monomial and knowledge of the source vertex. On the other
hand, two different labeled walks that are not paths can have the same
monomial, for example,
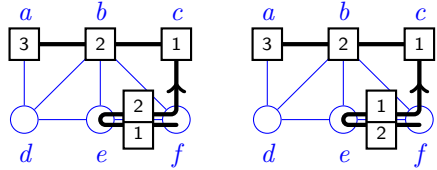
\[
\begin{tikzpicture}
  \begin{scope}[draw=blue]\exgraph\end{scope}
 \draw[walk]
 ([yshift=-2pt]F.center)--node[yshift=-3pt,lbl]{1} 
 ([yshift=-2pt]E.center)--
 ([yshift=2pt]E.center)--node[yshift=3pt,lbl]{2}
 ([yshift=2pt]F.center)--
 (C.center)--node[at start, lbl]{1}
 (B.center)--node[at start, lbl]{2} node[at end, lbl]{3}
 (A.center);
\end{tikzpicture}
\qquad
\begin{tikzpicture}
  \begin{scope}[draw=blue]\exgraph\end{scope}
 \draw[walk]
 ([yshift=-2pt]F.center)--node[yshift=-3pt,lbl]{2} 
 ([yshift=-2pt]E.center)--
 ([yshift=2pt]E.center)--node[yshift=3pt,lbl]{1}
 ([yshift=2pt]F.center)--
 (C.center)--node[at start, lbl]{1}
 (B.center)--node[at start, lbl]{2} node[at end, lbl]{3}
 (A.center);
\end{tikzpicture}
\]
are both associated with
\[x_{\edge{ab}}\cdot x_{\edge{bc}}\cdot x_{\edge{cf}}\cdot
x_{\edge{ef}}^2\cdot y_{a,3} \cdot
y_{b,2}\cdot y_{c,1} \cdot z_{\edge{ef},1}\cdot
z_{\edge{ef},2}\,.\] 

\begin{figure}
\begin{tabular}{c@{\hspace{2.5cm}}c}
$\myR_1$: label transposition &
$\myR_2$: labeled reversal \\
\def\arrowposition{.9}
\begin{tikzpicture}
 \begin{scope}[draw=blue]\exgraph\end{scope}
 \draw[walk]
 {[rounded corners=15pt]
   (C.center)--node[at start,lbl]{3} node[at end,xshift=5pt,lbl]{2}
 (B.center)}--
 (F.center)--node[lbl]{1}
 (E.center)--node[lbl]{2}
 (D.center)--
 (B.center) node[at end,xshift=-5pt,lbl]{1};
\end{tikzpicture}
&
\def\arrowposition{1}
\begin{tikzpicture}
  \begin{scope}[draw=blue]\exgraph\end{scope}
 \draw[walk]
 (C.center)-- node[at start, lbl]{3}
 (F.center)--
 (B.center)-- node[at start, lbl]{2} node[at end,lbl]{1}
 (A.center)-- 
 (D.center)-- node[lbl]{1}
 (E.center)-- node[lbl]{2}
 ([xshift=-5pt]F.center);
\end{tikzpicture}  
\\
\def\arrowposition{.9}
\begin{tikzpicture}
 \begin{scope}[draw=blue]\exgraph\end{scope}
 \draw[walk]
{[rounded corners=15pt] 
  (C.center)--node[at start,lbl]{3} node[at end,shift={(5pt,-0pt)},lbl]{1}
 (B.center)}--
 (F.center)--node[lbl]{1}
 (E.center)--node[lbl]{2}
 (D.center)--
 ([shift={(-2pt,-2pt)}]B.center) node[at end,shift={(-3pt,2pt)},lbl]{2};
\end{tikzpicture}
&
\def\arrowposition{1}
\begin{tikzpicture}
  \begin{scope}[draw=blue]\exgraph\end{scope}
 \draw[walk]
 (C.center)-- node[at start, lbl]{3}
 (F.center)--node[lbl]{2}
 (E.center)--node[lbl]{1}
 (D.center)--
 (A.center)-- node[at end, lbl]{2} node[at start,lbl]{1}
 (B.center)--
 ([shift={(-4pt,4pt)}]F.center);
\end{tikzpicture}  
\end{tabular}

\caption{\label{fig: pairing}\small Representative examples of the pairing of labeled
  walks. Left: if the walk's first closed subwalk starts at a vertex
  in $V_1=\{a,b,c\}$ (here, vertex $c$), then
  the labels at this vertex are transposed, while the walk remains
  unchanged.  Right: if the walk's first closed subwalk starts at a
  vertex in $V_2=\{d,e,f\}$ (here, vertex $f$),
  then the subwalk and its labels are reversed.}
\end{figure}
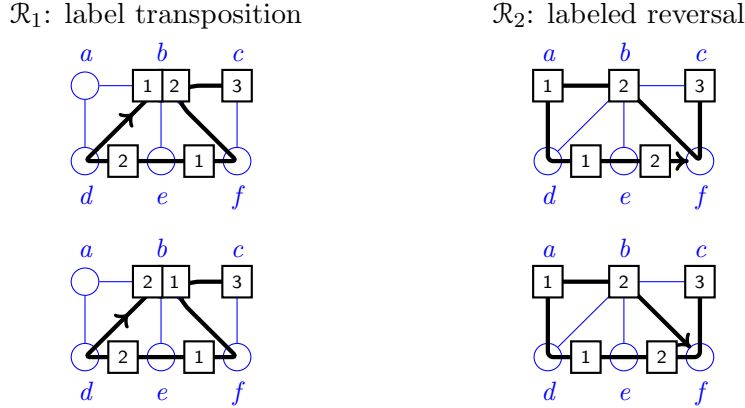

In fact, we will set up a pairing such that every labeled non-path has
exactly one such partner. In particular, their (identical) monomials
will cancel each other when added in a field of characteristic 2, and
only the monomials corresponding to labeled paths will
remain. Representative examples of this pairing are given in
Figure~\ref{fig: pairing}. A good part of our exposition is devoted to
a very careful description of this pairing; to appreciate why such
caution is necessary, note that walks like
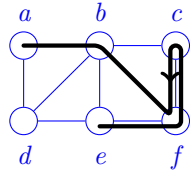
\[
\begin{tikzpicture}
  \begin{scope}[draw=blue]\exgraph\end{scope}
 \draw[walk]
 ([yshift=-2pt]E.center)--
 ([shift={(2pt,-2pt)}]F.center)--
 ([xshift=2pt]C.center)--
 ([xshift=-2pt]C.center)--
 ([shift={(-2pt,2pt)}]F.center)--
 (B.center)--
 (A.center);
\end{tikzpicture}
\]
are not correctly handled by our set-up and indeed will require an
exception in the definition.

\medskip In summary, there are four key ingredients.  First, the
labeled $k$-walks that are paths will be associated with monomials
that have distinct variable supports.  Second, the monomials
associated with {\em bijectively} labeled $k$-walks that are not paths
will cancel over a field of characteristic 2, establish by a pairing
argument.  Third, an inclusion--exclusion sieve will be used to cancel
all walks that are not bijectively labeled.  The sieve requires at
most $3k/4$ labels (with high probability) if we are careful about the
walks we consider.  Fourth, the polynomial of $k$-walks that {\em
  avoid} a given set of labels can be evaluated in time polynomial in
$n$.

\subsection{Preliminaries on strings}

From a technical perspective it
will be convenient to view a walk in a graph as a string.
To this end, let us review some basic terminology on strings.

Let $A$ be a set whose elements we view as the symbols
of an alphabet.
A {\em string} of {\em length} $\ell$ over $A$ 
is a sequence $S=s_1s_2\cdots s_\ell$ with $s_i\in A$ for 
each $i=1,2,\ldots,\ell$. We say that $s_i$ is the symbol
at {\em position} $i$ of the string.
The {\em reverse} of a string $S=s_1s_2\cdots s_\ell$
is $\overleftarrow S=s_\ell s_{\ell-1}\cdots s_1$.
The {\em concatenation} of two strings 
$S=s_1s_2\cdots s_\ell$ and $T=t_1t_2\cdots t_k$ is
$ST=s_1s_2\cdots s_\ell t_1t_2\cdots t_k$.
A string $T=t_1t_2\ldots t_k$ is a {\em substring}
of a string $S=s_1s_2\cdots s_\ell$ if there
exists a $j=1,2,\ldots,\ell-k+1$ such that 
$t_i=s_{i+j-1}$ for all $i=1,2,\ldots,k$.
A {\em palindrome} is a string that is identical 
to its reverse and has length at least $2$.
For $A_1,A_2,\ldots,A_\ell\subseteq A$, we say that
a string $s_1s_2\cdots s_\ell$ is an 
$A_1A_2\cdots A_\ell$-{\em string}
if $s_i\in A_i$ holds for every $i=1,2,\ldots,\ell$.

\subsection{Walks in graphs}
\label{sec: walks}
We assume that all graphs are undirected and contain 
neither loops nor parallel edges.
For a graph $G$, we denote the vertex set of $G$ by $V=V(G)$,
and the edge set of $G$ by $E=E(G)$.
For convenience, we assume that $V$ and $E$ are disjoint sets.

A $k$-{\em walk} in $G$ is a string of length $2k-1$ 
such that 
\begin{itemize}
\item[(a)] each odd position contains a vertex of $G$;
\item[(b)] each even position contains an edge of $G$; and
\item[(c)] for every $i=1,2,\ldots,k-1$,
the edge at position $2i$ joins in $G$ the vertices at positions 
$2i-1$ and $2i+1$.
\end{itemize}
The first and last positions of a walk are the {\em ends} 
of the walk.

A walk is a {\em path} if each vertex of $G$ appears in 
at most one position of the walk. 
A $k$-walk is {\em closed} if its ends are identical 
and $k\geq 2$. A walk that is not a path always contains 
at least one closed subwalk.

Let $W$ be a walk.
A {\em subwalk} of $W$ is a substring of $W$
that is in itself a walk. Put otherwise, 
a subwalk of $W$ is a substring with ends 
at odd positions of $W$.

\subsection{Representing strings as sets}

Let $a_1a_2\cdots a_\ell$ be a string over an alphabet $A$.
It will be convenient to view a string as a set consisting
of pairs $(a,i)\in A\times\{1,2,\ldots,\ell\}$, where the pair
$(a,i)$ indicates that the symbol $a$ occurs at position $i$,
that is, $a_i=a$.

For a subset $B\subseteq A$ and a string $a_1a_2\cdots a_\ell$, 
introduce the notation
\[
B\{a_1a_2\cdots a_\ell\}=
\{(a_i,i):a_i\in B\}\subseteq A\times\{1,2,\ldots,\ell\}.
\]
In particular, we can recover the string 
$a_1a_2\cdots a_\ell$ from the set $A\{a_1a_2\cdots a_\ell\}$.

\subsection{Admissible walks}
\label{sec: admissible}
To reduce the number of labels in the sieve, 
we will focus on a somewhat technical subset of 
$k$-walks that we will call ``admissible'' walks. 

Let $G$ be a graph with vertex set $V$
and let $s$ be a fixed vertex of $G$.

Partition the vertex set into two disjoint sets $V=V_1\cup V_2$. 
Denote by $E_1$ the set of edges of $G$ with both ends in $V_1$.
Denote by $E_2$ the set of edges of $G$ with both ends in $V_2$.
Let $k,k_1,\ell_2$ be nonnegative integers.

Let us say that a $k$-walk $W$ in $G$ is {\em admissible} if 
\begin{itemize}
\item[(a)] $W$ starts at $s$;
\item[(b)] $|V_1\{W\}|=k_1$; 
\item[(c)] $|E_2\{W\}|=\ell_2$; and 
\item[(d)] $W$ is $V_2EV_1EV_2$-palindromeless.
\end{itemize}
Here the term ``palindromeless'' refers to the property 
that a string has no palindrome as a substring. 
By $V_2EV_1EV_2$-palindromeless we refer to the lack of
palindromes that are also $V_2EV_1EV_2$-strings. 
Observe that paths are palindromeless and hence 
$V_2EV_1EV_2$-palindromeless.

\begin{figure}
\[
\vcenter{\hbox{\begin{tikzpicture}
\begin{scope}[draw=blue]\exgraph\end{scope}
\begin{scope}[font=\it,transparent]
\draw (A)--(B) node [midway,opaque] {A};
\draw (B)--(C) node [midway,opaque] {B};
\draw (A)--(D) node [midway,opaque] {C};
\draw (B)--(D) node [midway,opaque] {D};
\draw (B)--(E) node [midway,opaque] {E};
\draw (B)--(F) node [midway,opaque] {F};
\draw (C)--(F) node [midway,opaque] {G};
\draw (D)--(E) node [midway,opaque] {H};
\draw (F)--(E) node [midway,opaque] {I};
\end{scope}
\end{tikzpicture}}}
\qquad
W_1=\vcenter{\hbox{
\begin{tikzpicture}
\def\arrowposition{1}
\begin{scope}[draw=blue]\exgraph\end{scope}
\draw[walk] (A.center)--(D.center)--(B.center)--(E.center)--(F.center)--(C.center);
\end{tikzpicture}}}
\qquad
W_2=\vcenter{\hbox{
\begin{tikzpicture}
\begin{scope}[draw=blue]\exgraph\end{scope}
 \draw[walk]
 ([yshift=-2pt]E.center)--
 ([shift={(2pt,-2pt)}]F.center)--
 ([xshift=2pt]C.center)--
 ([xshift=-2pt]C.center)--
 ([shift={(-2pt,2pt)}]F.center)--
 (B.center)--
 (A.center);
\end{tikzpicture}}}
\]
\caption{\label{fig: terminology}\small Terminology and notation for
  walks introduced in \S\ref{sec: walks}--\ref{sec: admissible}.  The
  6-vertex walk $W_1$ is encoded by the 11-letter string {\it
    aCdDbEeIjGc}.  With $V_1=\{a,b,c\}$ we have $E_2=\{H,I\}$ and the
  ``projections'' $V_1\{W_1\}=\{(a,1),(b,5),(c,11)\}$ and
  $E_2\{W_1\}=\{(I,8)\}$.  The walk $W_1$ is admissible with the
  following parameters: starting vertex $s=a$, number of vertices
  $k=6$, number of $V_1$-vertices $k_1=3$, and number of $E_2$-edges
  $\ell_2=1$.  The walk $W_2= \text{\it eIfGcGfFbAa}$ is not
  admissible, because it contains the $V_2EV_1EV_2$-palindrome {\it
    fGcGf}. For readability, in all other examples we denote edges by
  node pairs, for instance writing $\edge{ef}$\/ instead of $I$, with
  the understanding that $\edge{ef}$ is viewed as a single symbol for
  purposes of indexing $W$.}
\end{figure}
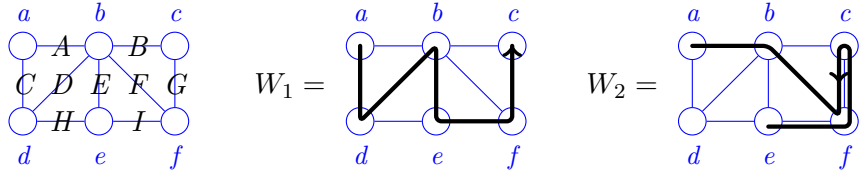

\subsection{Random projection}
\label{sect:projection}

For a fixed ordered partition $(V_1,V_2)$, 
every $k$-path $P$ in $G$ that starts at $s$ is 
admissible for some parameters $k_1,\ell_2$. 
Conversely, for fixed parameters $k_1,\ell_2$
and a fixed $k$-path $P$ that starts at $s$,
if we select $(V_1,V_2)$ uniformly at random,
then $P$ is admissible with probability given 
by the following lemma.

\begin{Lem}[Admissibility]
\label{lem:adm-prob}
Let $k_1,\ell_2$ be nonnegative integers and 
let $P$ be a $k$-path in $G$.
For $(V_1,V_2)$ selected uniformly at random,
we have
\[
\Pr\biggl(\text{$|V_1\{P\}|=k_1$ and\/ $|E_2\{P\}|=\ell_2$}\biggr)
=2^{-k}\binom{k_1+1}{k-k_1-\ell_2}\binom{k-k_1-1}{\ell_2}.
\]
\end{Lem}
\begin{Proof}
There are $2^k$ strings of length $k$ over the alphabet $\{1,2\}$. 
The probability in question is exactly the fraction of such
strings that have exactly $k_1$ $1$-positions and exactly 
$\ell_2$ $22$-substrings. There are exactly
$k_1+1$ positions where to interleave the $k_1$ 1s with 
substrings of 2s. Each such substring of length $j$ contributes 
exactly $j-1$ $22$-substrings. The total number of 2s is $k-k_1$, 
so there must be $k-k_1-\ell_2$ substrings of 2s.
The positions where the substrings interleave the 1s
are allocated by the first binomial coefficient. 
It remains to allocate the lengths of the strings.
The total length is $k-k_1$, and each 
of the $k-k_1-\ell_2$ strings must have length at least 1. 
Thus there are $k-k_1-(k-k_1-\ell_2)=\ell_2$ 
free 2s to allocate to $k-k_1-\ell_2$ distinct bins. 
The second binomial coefficient carries out this allocation.
\end{Proof}

In particular, a fixed $k$-path starting at $s$ is admissible 
with positive probability if and only if
either $k_1=k$ and $\ell_2=0$ 
or $k_1<k$ and $k_1+\ell_2\leq k-1\leq 2k_1+\ell_2$.

Let us now derive an asymptotic approximation for the probability
in Lemma~\ref{lem:adm-prob}.
We employ the following variant of Stirling's formula due to
Robbins~\cite{Robbins1955}. For all $j=1,2,\ldots$ it holds that
\begin{equation}
\label{eq:stirling}
j!=\sqrt{2\pi j}\biggl(\frac{j}{e}\biggr)^je^{\epsilon_j}\quad\text{where}\quad\frac{1}{12j+1}<\epsilon_j<\frac{1}{12j}\,.
\end{equation}
Let us abbreviate 
\[
\bangle{a}{b}=\biggl(\frac{b}{a}\biggr)^{-b}\biggl(1-\frac{b}{a}\biggr)^{-a+b}\,.
\]
From Stirling's formula \eqref{eq:stirling}
it follows that $\binom{a}{b}=\Theta^*\bigl(\bangle{a}{b}\bigr)$
holds uniformly for all $0<b<a\leq n$. 
We can thus approximate the probability in Lemma~\ref{lem:adm-prob}, 
uniformly for $0<\ell_2,k_1<k$ such that 
$k_1+\ell_2\leq k-1\leq 2k_1+\ell_2$, with
\begin{equation}
\label{eq:walk-adm-prob-bound}
\begin{split}
\Pr&\biggl(\text{$|V_1\{P\}|=k_1$ and\/ $|E_2\{P\}|=\ell_2$}\biggr)=\\
&\qquad\qquad=
\Theta^*\biggl(2^{-k}
\bangle{k_1}{k-k_1-\ell_2}\!
\bangle{k-k_1}{\ell_2}\!
\biggr)\,.
\end{split}
\end{equation}

\subsection{Labeled admissible walks}

The following labeling scheme for admissible walks serves 
two purposes. First, labeling enables us to ``decouple'' 
the sieve from the graphical domain (that is, vertices and edges) 
into a set of abstract labels whose number depends only 
on the parameters $k_1,\ell_2$ and not on the size 
of the graph. Second, the labeling facilitates cancellation 
of non-paths in the sieve. 

Let $K_1$ be a set of $k_1$ labels. 
Let $L_2$ be a set of $\ell_2$ labels.
For example, let $K_1=\{1,2,\ldots,k_1\}$ and 
$L_2=\{1,2,\ldots,\ell_2\}$.

Let $W$ be an admissible walk.
Let $\kappa_1: V_1\{W\}\rightarrow K_1$ 
and $\lambda_2: E_2\{W\}\rightarrow L_2$ 
be arbitrary functions.
The three-tuple $(W,\kappa_1,\lambda_2)$ is a {\em labeled} admissible walk.
Intuitively, each position in $W$ that contains a vertex in $V_1$
gets assigned a label in $K_1$ by $\kappa_1$. Similarly, each
position in $W$ that contains an edge in $E_2$ gets assigned
a label in $L_2$ by $\lambda_2$.
Let us say that the labeling is {\em bijective} if
both $\kappa_1$ and $\lambda_2$ are bijections.

\medskip
\noindent{\em Example.}
Consider two labelings of the same walk $W$,
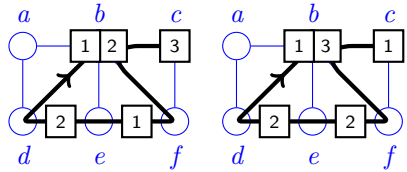
\[\def\arrowposition{.9}
\begin{tikzpicture}
 \begin{scope}[draw=blue]\exgraph\end{scope}
 \draw[walk]
 {[rounded corners=15pt]
   (C.center)--node[at start,lbl]{3} node[at end,xshift=5pt,lbl]{2}
 (B.center)}--
 (F.center)--node[lbl]{1}
 (E.center)--node[lbl]{2}
 (D.center)--
 (B.center) node[at end,xshift=-5pt,lbl]{1};
\end{tikzpicture}
\quad
\begin{tikzpicture}
 \begin{scope}[draw=blue]\exgraph\end{scope}
 \draw[walk]
 {[rounded corners=15pt]
   (C.center)--node[at start,lbl]{1} node[at end,xshift=5pt,lbl]{3}
 (B.center)}--
 (F.center)--node[lbl]{2}
 (E.center)--node[lbl]{2}
 (D.center)--
 (B.center) node[at end,xshift=-5pt,lbl]{1};
\end{tikzpicture}
\]
The walk is admissible with parameters $s=\mathrm c$, $k=6$, $k_1=3$,
and $\ell_2=2$.  Both labelings associate a label with each position
of $W$ that contains a symbol from $V_1=\{a,b,c\}$ or
$E_2=\{\edge{de},\edge{ef}\,\}$. We have
$V_1\{W\}=\{(b,3),(b,11),(c,1)\}$, that is, there are three
occurrences of symbols from $V_1$ in $W$; in particular the symbol $b$
occurs at the 3rd and the 11th position. Similarly, we have
$E_2\{W\}=\{(\edge{ef},6),(\edge{de},8)\}$.  The labeling on the left
is
\[
\lambda_1( b,3)=2,\;
\lambda_1( b,11)=1,\;
\lambda_1( b,1)=3,\;
\kappa_2(\edge{ef},6)=2,\;
\kappa_2(\edge{de},8)=1.
\]
We observe that this labeling is bijective since
$\lambda_1$ and $\kappa_2$ are bijections.

The labeling on the right is \[
\lambda_1( b,3)=3,\;
\lambda_1( b,11)=
\lambda_1( c,1)=1,\;
\kappa_2(\edge{ef},6)=
\kappa_2(\edge{de},8)=2,
\]
and not bijective. In fact, $\lambda_1$ avoids the label 2, and
$\kappa_2$ avoids the label 1.

\subsection{Fingerprinting and identifiability}

We associate with each labeled admissible walk
an algebraic object (or ``fingerprint'') that we use 
to represent the labeled admissible walk in sieving. 
Here it is important to observe that while we are
careful to design the fingerprint so that each labeled
path has a unique fingerprint, the fingerprints of
labeled non-paths are by design {\em not} unique---%
we will explicitly take advantage of this property 
when canceling labeled non-paths 
in \S\ref{sect:non-paths-cancel}.

The sieve operates over a multivariate polynomial ring with 
the coefficient field $\F_{2^b}$ (the finite field of order $2^b$)
and the following indeterminates. 
Introduce one indeterminate $x_e$ for each edge $e\in E$.
Introduce one indeterminate 
$y_{v,i}$ for each pair $(v,i)\in V_1\times K_1$.
Introduce one indeterminate 
$z_{e,i}$ for each pair $(e,i)\in E_2\times L_2$.

Let $(W,\kappa_1,\lambda_2)$ be a labeled admissible walk.
Associate with $(W,\kappa_1,\lambda_2)$ the 
{\em monomial} ({\em fingerprint}\/)
\[
m(W,\kappa_1,\lambda_2)=
\prod_{(e,j)\in E\{W\}} x_e
\prod_{(v,i)\in V_1\{W\}} y_{v,\kappa_1(v,i)}
\prod_{(e,i)\in E_2\{W\}} z_{e,\lambda_2(e,i)}\,.
\]

The following lemma is immediate.

\begin{Lem}[Identifiability]
\label{lem:walk-identifiability}
The monomial $m(W,\kappa_1,\lambda_2)$ of a labeled
admissible walk $(W,\kappa_1,\lambda_2)$ uniquely
determines the edges and their multiplicities 
of occurrence in $W$. In particular, any path
is uniquely identified. Furthermore, if $W$ is 
a path and $\kappa_1,\lambda_2$ are bijections,
then $m(W,\kappa_1,\lambda_2)$ uniquely
identifies $(W,\kappa_1,\lambda_2)$.
\end{Lem}

\medskip
\noindent{\em Example.}
We presented some example monomials already in \S\ref{sec: path
  overview}. We can also consider a bijectively labeled walk that
repeats an edge in $E_2$:
\[
\vcenter{\hbox{ \begin{tikzpicture}
 \begin{scope}[draw=blue]\exgraph\end{scope}
 \draw[walk]
 ([shift={(0pt,-2pt)}]F.center)-- node[lbl,yshift=-3pt]{1}
([shift={(2pt,-2pt)}]E.center)-- node[lbl, at end, shift={(5pt,3pt)}]{2}
 ([shift={(2pt,-2pt)}]B.center)--
 ([shift={(-2pt,2pt)}]F.center)-- node[lbl,yshift=3pt]{2}
 ([shift={(-2pt,2pt)}]E.center)-- node[lbl, at end,shift={(-4pt,1pt)}]{1}
 ([shift={(-2pt,2pt)}]B.center);
\end{tikzpicture}}}
\qquad
x_{\edge{be}}^2\cdot  x_{\edge{bf}}\cdot x_{\edge{ef}}^2 \cdot y_{
  b,1}\cdot y_{b,2}
\cdot z_{\edge{ef},1}
\cdot z_{\edge{ef},2}\,,
\]
and a non-bijectively labeled walk,
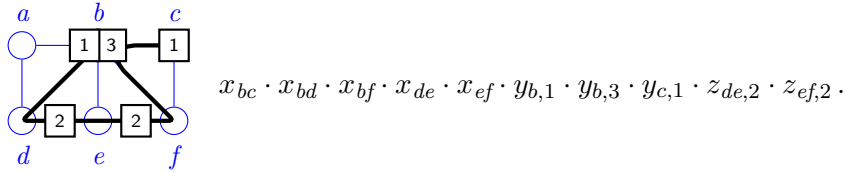
\[\vcenter{\hbox{\begin{tikzpicture}
 \begin{scope}[draw=blue]\exgraph\end{scope}
 \draw[walk]
 {[rounded corners=15pt]
   (C.center)--node[at start,lbl]{1} node[at end,xshift=5pt,lbl]{3}
 (B.center)}--
 (F.center)--node[lbl]{2}
 (E.center)--node[lbl]{2}
 (D.center)--
 (B.center) node[at end,xshift=-5pt,lbl]{1};
\end{tikzpicture}}}
\quad
x_{\edge{bc}}\cdot  x_{\edge{bd}}\cdot x_{\edge{bf}}\cdot x_{\edge{de}}\cdot x_{\edge{ef}} \cdot y_{b,1} \cdot y_{b,3} \cdot y_{c,1}
\cdot z_{\edge{de},2}
\cdot z_{\edge{ef},2}\,.
 \]

 In all these examples observe that if the walk is a path, we can
 reconstruct it from the $x$-variables and
 knowledge of the start vertex. Because a path has neither repeated
 vertices nor edges, the $y$- and $z$-variables in the monomial enable
 us to reconstruct the labeling.

\subsection{Sieving for bijective labelings}

Let us denote by $\myL$ the set of all 
labeled admissible walks. For $I_1\subseteq K_1$
and $J_2\subseteq L_2$, denote by $\myL[I_1,J_2]$ 
the set of all labeled admissible walks 
that {\em avoid} the labels in $I_1$ and $J_2$.
Let us denote by $\myB$ the set of all 
bijectively labeled admissible walks.
 
By the principle of inclusion--exclusion, we have
\begin{equation}
\label{eq:labeling-sieve}
\sum_{(W,\kappa_1,\lambda_2)\in\myB}\!\!\!\!\!\! m(W,\kappa_1,\lambda_2)=
\sum_{I_1\subseteq K_1}
\sum_{J_2\subseteq L_2}
(-1)^{|I_1|+|J_2|}
\!\!\!\!\!\!\!\!\!\!\!\!\sum_{(W,\kappa_1,\lambda_2)\in\myL[I_1,J_2]}\!\!\!\!\!\!\!\!\!\!\!\!m(W,\kappa_1,\lambda_2)\,.
\end{equation}

\subsection{Bijectively labeled non-path fingerprints cancel}
\label{sect:non-paths-cancel}

Let us partition $\myB$ into $\myB=\myP\cup\myR$, where
$\myP$ consists of bijectively labeled admissible paths, 
and $\myR$ consists of bijectively labeled admissible non-paths.
Accordingly, 
the left-hand side of \eqref{eq:labeling-sieve} splits into
\[
\sum_{(W,\kappa_1,\lambda_2)\in\myB}\!\!\!\!\!\! m(W,\kappa_1,\lambda_2)=
\!\!\!\!\!\!\sum_{(W,\kappa_1,\lambda_2)\in\myP}\!\!\!\!\!\! m(W,\kappa_1,\lambda_2)+
\!\!\!\!\!\!\sum_{(W,\kappa_1,\lambda_2)\in\myR}\!\!\!\!\!\! m(W,\kappa_1,\lambda_2)\,.
\]
We show that the rightmost sum vanishes. 
To this end, let us first recall that an {\em involution}
is a permutation that is its own inverse. 
We claim that it suffices to construct a fixed-point-free involution
$\phi:\myR\rightarrow\myR$ 
with $m(W,\kappa_1,\lambda_2)=m(\phi(W,\kappa_1,\lambda_2))$
for all $(W,\kappa_1,\lambda_2)\in\myR$,
Indeed, introduce an arbitrary total order to $\myR$
and observe that in characteristic 2, 
we have
\[
\begin{split}
\sum_{(W,\kappa_1,\lambda_2)\in\myR}m(W,\kappa_1,\lambda_2)
&=
\!\!\!\sum_{\substack{(W,\kappa_1,\lambda_2)\in\myR\\(W,\kappa_1,\lambda_2)<\phi(W,\kappa_1,\lambda_2)}}\!\!\!\!\!\!m(W,\kappa_1,\lambda_2)+m(\phi(W,\kappa_1,\lambda_2))\\
&=
\!\!\!\sum_{\substack{(W,\kappa_1,\lambda_2)\in\myR\\(W,\kappa_1,\lambda_2)<\phi(W,\kappa_1,\lambda_2)}}\!\!\!\!\!\!2m(W,\kappa_1,\lambda_2)\quad=\quad0\,.
\end{split}
\]

To construct a fixed-point-free
involution $\phi:\myR\rightarrow\myR$
with $m(W,\kappa_1,\lambda_2)=m(\phi(W,\kappa_1,\lambda_2))$
for all $(W,\kappa_1,\lambda_2)\in\myR$,
we observe that every walk $W$ that 
is not a path contains at least one closed subwalk. 
In particular, $W$ contains a {\em first} closed subwalk, 
that is, the closed subwalk $C$ with the property that $C$ is 
the unique closed subwalk in the prefix $SC$ of $W=SCT$.

We denote the first closed subwalk of $W$ by $C(W)$
and by $c(W)$ the first (and hence also the last) vertex of $C(W)$.

Let us partition $\myR$ into two disjoint sets, $\myR_1$ and $\myR_2$,
where
\begin{align*}
\myR_1&=\{(W,\kappa_1,\lambda_2)\in\myR:c(W)\in V_1\}\,,\\
\myR_2&=\{(W,\kappa_1,\lambda_2)\in\myR:c(W)\in V_2\}\,.
\end{align*}
We proceed to construct the pairing $\phi$ on these two sets. See
Figure~\ref{fig: pairing} for examples.

\subsection{The pairing on $\myR_1$ -- label transposition}
\label{sect:pairing-label-trans}

Select an arbitrary $(W,\kappa_1,\lambda_2)\in\myR_1$.
Let $j$ and $\ell$ be the positions of $W$ that contain the symbol
$c(W)$ and constitute the ends of $C(W)$. For brevity, 
let us write $c$ for $c(W)$.
Because $c\in V_1$, we have $(c,j),(c,\ell)\in V_1\{W\}$.
Define $\kappa_1'$ to be identical
to $\kappa_1$ except that 
\[
\kappa_1'(c,j)=\kappa_1(c,\ell)\,,\qquad
\kappa_1'(c,\ell)=\kappa_1(c,j)\,.
\]
Observe that $\kappa_1(c,j)\neq\kappa_1(c,\ell)$ 
because $(W,\kappa_1,\lambda_2)$ is bijectively labeled.
Thus, $\kappa_1'\neq\kappa_1$ and 
$(W,\kappa_1',\lambda_2)\in\myR_1$.
Furthermore, we have $m(W,\kappa_1,\lambda_2)=\allowbreak  m(W,\kappa_1',\lambda_2)$.
Thus, we can set $\phi(W,\kappa_1,\lambda_2)=(W,\kappa_1',\lambda_2)$
to obtain the desired fixed-point-free involution on $\myR_1$.
Indeed, 
$\phi(W,\kappa_1,\lambda_2)=(W,\kappa_1',\lambda_2)\neq(W,\kappa_1,\lambda_2)$
and $\phi^2(W,\kappa_1,\lambda_2)=(W,\kappa_1,\lambda_2)$.

\subsection{The pairing on $\myR_2$ -- labeled reversal of first closed subwalk}

Select an arbitrary $(W,\kappa_1,\lambda_2)\in\myR_2$.
Let $C=C(W)$ and let $S,T$ be strings such that 
\[
W=SCT\,.
\]
Let us define the string $W'$ by reversing $C$ in $W$, 
that is,
\[
W'=S\overleftarrow{C}T\,.
\]
We observe that 
the strings $C$ and $\overleftarrow C$ have identical ends
because $C$ is a closed walk, implying that $W'$ is a walk in $G$.
We also observe that $W'$ is admissible.
Indeed, because $c(W')=c(W)\in V_2$, any $V_2EV_1EV_2$-palindrome 
$ueveu$ in $W'$ can either (a) occur as a subwalk of 
$\overleftarrow{C}$ in $W'$, or (b) have at most one position
in $W'$ common with $\overleftarrow{C}$. But in both cases
we have that $ueveu$ occurs in $W$, which is a contradiction 
since $W$ is admissible. Thus, $W'$ is admissible.

We observe that $W=W'$ if and only if $C$ is a palindrome.
Furthermore, if we reverse $C(W')=\overleftarrow{C}$ in $W'$,
we obtain back $W$. That is, $W''=W$. 

In terms of string positions, we can characterize
the reversal $W\mapsto W'$ using the following permutation
of positions. Let $j$ and $\ell$ be the positions of $W$ that 
constitute the ends of $C$. Define the permutation 
$\rho:\{1,2,\ldots,k\}\rightarrow\{1,2,\ldots,k\}$
by
\[
\rho(i)=
\begin{cases}
i & \text{if $i<j$ or $i>\ell$; and}\\
\ell-i+j & \text{if $j\leq i\leq \ell$.}
\end{cases}
\]
Let us denote the symbol at the $i$th position of $W$ by $w_i$.
The reversal $W\mapsto W'$ can now be characterized by
observing that $w'_{\rho(i)}=w_i$ holds for each $i=1,2,\ldots,k$.

We now introduce a labeling $\kappa_1',\lambda_2'$ of $W'$
using the labeling $\kappa_1,\lambda_2$ of $W$.
In particular, let us label $W'$ so that each position 
of $W'$ is labeled using the label of the $\rho$-corresponding 
position in $W$, if any. In precise terms, 
using the labeling $\kappa_1: V_1\{W\}\rightarrow K_1$,
define the labeling $\kappa_1': V_1\{W'\}\rightarrow K_1$
for each $(w'_{\rho(i)},\rho(i))\in V_1\{W'\}$
by setting $\kappa_1'(w'_{\rho(i)},\rho(i))=\kappa_1(w_i,i)$.
Similarly, using $\lambda_2: E_2\{W\}\rightarrow L_2$,
define $\lambda_2': E_2\{W'\}\rightarrow L_2$ by
setting $\lambda_2'(w'_{\rho(i)},\rho(i))=\lambda_2(w_i,i)$
for each  $(w'_{\rho(i)},\rho(i))\in E_2\{W'\}$.

Now set
$\phi(W,\kappa_1,\lambda_2)=(W',\kappa_1',\lambda_2')$
and observe that $\phi(W,\kappa_1,\lambda_2)\in\myR_2$,
$\phi^2(W,\kappa_1,\lambda_2)=(W,\kappa_1,\lambda_2)$, and
$m(W,\kappa_1,\lambda_2)=m(\phi(W,\kappa_1,\lambda_2))$.

What is not immediate, however, is that 
$\phi(W,\kappa_1,\lambda_2)\neq(W,\kappa_1,\lambda_2)$.
There are two cases to consider, depending on $C$.

In the first case, $C$ is not a palindrome, that is, 
$C\neq \overleftarrow C$. Thus, $W'\neq W$ and 
hence $(W',\kappa_1',\lambda_2')\neq(W,\kappa_1,\lambda_2)$.

In the second case, $C$ is a palindrome. Since $C$
is a closed walk, the string $C$ has odd length
at least 3. In particular, the length 3 (that is, a palindrome
of the form $ueu$ with $u\in V_2$ and $e\in E$) cannot occur 
because $G$ has no loop edges.
For palindromes of length 5, the only possibility is that $C$ is a
$V_2E_2V_2E_2V_2$-palindrome. 
Indeed, $C$ can neither be a $V_1EV_1EV_1$-palindrome 
nor a $V_1EV_2EV_1$-palindrome because $c(W)\in V_2$. 
Furthermore, $C$ cannot be a $V_2EV_1EV_2$-palindrome
because such palindromes by definition do not occur in the admissible $W$.
Thus, for length 5 the only possibility is a 
$V_2EV_2EV_2$-palindrome, that is, a $V_2E_2V_2E_2V_2$-palindrome.
Such a palindrome contains two occurrences of an edge in $E_2$
that are in $\rho$-corresponding positions.
These occurrences get different labels under $\lambda_2$ and $\lambda_2'$. 
Thus, $(W',\kappa_1',\lambda_2')\neq(W,\kappa_1,\lambda_2)$.
Finally, we observe that $C$ cannot have length more than 5,
because a palindrome of length 7 or more must include a palindrome
of length 5, which would contradict the assumption that $C$
is the first closed subwalk in $W$. 

\subsection{The algorithm}
\label{sec: p-packings of q-sets algorithm}

First, we recall the following result:
\begin{Lem}[DeMillo--Lipton--Schwartz--Zippel \cite{DeMilloLipton1978,Schwartz1980}]
\label{lem:dlsz}
Let $p(x_1,x_2,\ldots,x_n)$ be a nonzero polynomial 
of total degree at most $d$ over the finite field\/ $\F_q$. 
Then, for $a_1,a_2,\ldots,a_n\in\F_q$ selected independently 
and uniformly at random,
\[
\Pr(p(a_1,a_2,\ldots,a_n)\neq 0)\geq 1-\frac{d}{q}.
\]
\end{Lem}

Let us assume the parameters $k,k_1,\ell_2$ have been fixed
so that $k_1+\ell_2\leq k-1\leq 2k_1+\ell_2$.
(We will set the precise values of $k_1,\ell_2$ in what follows.)
To decide the existence of a $k$-path starting at $s$, 
we repeat the following randomized procedure.

First, the procedure selects an ordered partition 
$(V_1,V_2)$ uniformly at random among all the $2^n$ such partitions. 
Lemma~\ref{lem:adm-prob} implies that a fixed $k$-path $P$ that 
starts at $s$ is admissible with positive probability. 

Next, the procedure makes use of Lemma~\ref{lem:dlsz} to witness
a nonzero evaluation of a multivariate generating function for
labeled admissible $k$-paths starting at $s$. In particular, 
from \eqref{eq:labeling-sieve} and \S\ref{sect:non-paths-cancel}
we have that 
\begin{equation}
\label{eq:path-poly}
\sum_{(W,\kappa_1,\lambda_2)\in\myP}\!\!\! m(W,\kappa_1,\lambda_2)
=
\sum_{I_1\subseteq K_1}
\sum_{J_2\subseteq L_2}
(-1)^{|I_1|+|J_2|}
\!\!\!\!\!\!\!\!\!\!\sum_{(W,\kappa_1,\lambda_2)\in\myL[I_1,J_2]}\!\!\!\!\!\!\!\!\!\!\!m(W,\kappa_1,\lambda_2)\,.
\end{equation}
The left-hand side of \eqref{eq:path-poly} is a multivariate
polynomial of degree at most $k-1+k_1+\ell_2$. 
It follows from Lemma~\ref{lem:walk-identifiability} that 
the polynomial is not identically zero if and only if $G$ has 
an admissible $k$-path starting at $s$.

It remains to evaluate the right-hand side of \eqref{eq:path-poly} for 
a random assignment of values in $\F_{2^b}$ to the indeterminates.
To this end, the procedure iterates over each 
$I_1\subseteq K_1$ and $J_2\subseteq L_2$ and employs 
dynamic programming to evaluate the rightmost sum 
in \eqref{eq:path-poly}.

Without loss of generality we can assume $k\geq 3$. 
For parameters $k,k_1,\ell_2$ and a string
$T=t_1t_2t_3t_4t_5$ over the alphabet $V\cup E$, 
our objective is to compute
\begin{equation}
\label{eq:walk-suffix}
M(k,k_1,\ell_2,T)
=
\sum_{\substack{(W,\kappa_1,\lambda_2)\in\myL[I_1,J_2]\\\text{$T$ is a suffix of $W$}}}m(W,\kappa_1,\lambda_2)\,.
\end{equation}
In particular, taking the sum over all $T$, we obtain the
rightmost sum in \eqref{eq:path-poly}.

The recursion for \eqref{eq:walk-suffix} is as follows.
For a logical proposition $P$, 
let us define $[P]$ to be $1$ if $P$ is true and $0$ otherwise. 
For $k>3$, we observe by induction on $k$ that
\begin{equation}
\label{eq:walk-recursion}
\begin{split}
M&(k,k_1,\ell_2,t_1t_2t_3t_4t_5)=\\
&=[\text{$t_1t_2t_3t_4t_5$ is not a $V_2EV_1EV_2$-palindrome}]\\
&\,\times\!\!\!\sum_{\substack{e\in E\\\text{$e=\{v,t_1\}$}}}\!\!\!
\bigl([v\notin V_1]+[v\in V_1]\!\!\!\sum_{j\in K_1\setminus I_1}\!\!\!y_{v,j}\bigr)
\bigl([e\notin E_2]+[e\in E_2]\!\!\!\sum_{j\in L_2\setminus J_2}\!\!\!z_{e,j}\bigr)\\
&\qquad\qquad\quad\qquad\times M(k-1,k_1-[v\in V_1],\ell_2-[e\in E_2],vet_1t_2t_3)\,.
\end{split}
\end{equation}
To set up the base cases for the recursion,
we observe that $M(k,k_1,\ell_2,T)$ can be computed
for all $0\leq k_1,\ell_2\leq k=3$ and all $T=t_1t_2t_3t_4t_5$ 
in time polynomial in $n$. Furthermore, $M(k,k_1,\ell_2,T)=0$ 
whenever $k_1>k$ or $\ell_2\geq k$ or $k_1<0$ or $\ell_2<0$.

Consequently, for any given assignment of values in $\F_{2^b}$
to the indeterminates $x_e$, $y_{v,\ell}$, and $z_{e,\ell}$, 
the procedure evaluates the right-hand side of \eqref{eq:path-poly} 
via \eqref{eq:walk-recursion} in $O(2^{k_1+\ell_2}k^3n^4)$ arithmetic 
operations over $\F_{2^b}$. 

Let us now complete the algorithm by optimizing the parameters 
for running time and $\Omega(1)$ probability of success.
Denoting the probability that a $k$-path $P$ starting at $s$
is admissible by $P(k,k_1,\ell_2)$, we have that in $r$ 
repetitions of the procedure at least one repetition finds
$P$ admissible with probability 
$1-(1-P(k,k_1,\ell_2))^r\geq 1-e^{-P(k,k_1,\ell_2)r}$.
Setting $r=\lceil 1/P(k,k_1,\ell_2)\rceil$ and $b=\lceil\log_2 6k\rceil$, 
it follows from Lemma~\ref{lem:dlsz} that any fixed $k$-path starting 
at $s$ in $G$ is witnessed 
with probability at least $(1-e^{-1})/2$ in time 
$O\bigl(2^{k_1+\ell_2}k^5n^4/P(k,k_1,\ell)\bigr)$. 
Setting $k_1=\lfloor\gamma_1k\rfloor$, $\ell_2=\lfloor\gamma_2k\rfloor$,
and employing \eqref{eq:walk-adm-prob-bound} to approximate
$P(k,k_1,\ell_2)$, we obtain $O^*(1.6569^k)$ time 
for $\gamma_1=0.5$ and $\gamma_2=0.207107$.

\section{A Determinant Sieve for $q$-Dimensional $p$-Packings}
\label{sec: matching}

This section establishes Theorem~\ref{thm: matching}.

\subsection{Prepackings and Edmonds's symbolic determinant}

Let us say that a subset $\myA\subseteq\myF$ 
is a {\em $j$-prepacking} if $|\myA|=j$ and the sets in $\myA$ 
are pairwise disjoint when projected to $U_1\cup U_2$.

Observe that each $A\in\myA$ in a $j$-prepacking
identifies both 
a unique $u_1(A)\in A\cap U_1$ and 
a unique $u_2(A)\in A\cap U_2$. 

For a bijection $\sigma:U_1\rightarrow U_2$,
let us say that a $j$-prepacking $\myA$ is {\em compatible} 
with $\sigma$ if for all $A\in\myA$ it holds that 
$\sigma(u_1(A))=u_2(A)$.
Note that each $j$-prepacking is compatible with at least one $\sigma$.

Edmonds~\cite{Edmonds1967} made the algorithmically seminal observation 
that the determinant of a symbolic $r\times r$ matrix 
$E=(e_{u_1,u_2})_{u_1\in U_1,u_2\in U_2}$ 
is a signed generating function over partitions of 
$U_1\cup U_2$ into $2$-subsets with exactly one element from $U_1$ 
and exactly one element from $U_2$. Indeed, identifying each
such partition with a bijection $\sigma:U_1\rightarrow U_2$, we have
\begin{equation}
\label{eq:det-formula}
\det E=\sum_{\substack{\sigma:U_1\rightarrow U_2\\\text{$\sigma$ bijective}}}
\sgn_{\tau}(\sigma)\prod_{u_1\in U_1}e_{u_1,\sigma(u_1)}\,,
\end{equation}
where the sign $\sgn_\tau(\sigma)$ is the sign
of the permutation $\sigma\tau$ for an
arbitrary fixed bijection $\tau:U_2\rightarrow U_1$.

Our strategy is to leverage Edmonds's observation
from the dimensions $U_1$ and $U_2$ into
$q$ dimensions $U_1,U_2,\ldots,U_q$ with sieving.
In particular, Edmonds's observation forces
the packing constraint in the first two
dimensions, which allows us to restrict
the sieve to the remaining $q-2$ dimensions.

\subsection{Fingerprinting and identifiability}

Consider a $j$-prepacking $\myA\subseteq\myF$.
The {\em domain} of the prepacking is the set
\begin{equation}
\label{eq:det-domain}
d(\myA)
=\{(u,A):u\in A\in\myA\}
\subseteq (U_3\cup U_4\cup\cdots\cup U_q)\times\myF\,.
\end{equation}
Observe that $|d(\myA)|=j(q-2)$.

Let $L$ be a set of $p(q-2)$ labels.
A {\em labeling} of $\myA$ 
is a pair $(\sigma,\lambda)$, 
where $\sigma:U_1\rightarrow U_2$ is a bijection compatible 
with $\myA$ and $\lambda:d(\myA)\rightarrow L$ is an arbitrary 
mapping. The labeling is {\em bijective}
if $\lambda$ is a bijection. 
We say that a triple $(\myA,\sigma,\lambda)$ is 
a {\em labeled} $j$-prepacking.

The sieve operates over a multivariate polynomial ring with 
the coefficient field $\F_{2^b}$ and the following indeterminates. 
Introduce the indeterminate $w$ for tracking the weight
$j$ of a $j$-prepacking. 
Associate with each  
$A\in\myF$ an indeterminate $x_A$.
Associate with each pair $(u_1,u_2)\in U_1\times U_2$
an indeterminate $y_{u_1,u_2}$.
Associate with each pair 
$(u,\ell)\in (U_3\cup U_4\cup\cdots\cup U_q)\times L$ 
an indeterminate $z_{u,\ell}$.

The {\em signed monomial} of a labeled $j$-prepacking 
$(\myA,\sigma,\lambda)$ is
\begin{equation}
\label{eq:det-monomial}
m(\myA,\sigma,\lambda)=
\sgn_\tau(\sigma)w^j\prod_{A\in\myA} x_{A}
\prod_{u_1\in U} y_{u_1,\sigma(u_1)}
\prod_{(u,A)\in d(\myA)} z_{u,\lambda(u,A)}\,.
\end{equation}

\begin{Lem}[Identifiability]
\label{lem:det-identifiability}
The monomial $m(\myA,\sigma,\lambda)$ uniquely determines 
both $\myA$ and $\sigma$. Furthermore, 
if $\myA$ is a $p$-packing and $\lambda$ is bijective, 
then $m(\myA,\sigma,\lambda)$ uniquely determines $\lambda$.
\end{Lem}

\subsection{Sieving for bijective labelings}

Denote by $\myL$ the set of all labeled $p$-prepackings.
For $J\subseteq L$, denote by $\myL[J]$ the subset of labeled
$p$-prepackings whose labeling {\em avoids} each label in $J$.
Denote by $\myB$ the set of all bijectively labeled $p$-prepackings.

By the principle of inclusion-exclusion,
\begin{equation}
\label{eq:det-sieve}
\sum_{(\myA,\sigma,\lambda)\in\myB} m(\myA,\sigma,\lambda)
=
\sum_{J\subseteq L}(-1)^{|J|}\sum_{(\myA,\sigma,\lambda)\in\myL[J]}
 m(\myA,\sigma,\lambda)\,.
\end{equation}

\subsection{Fingerprints of bijectively labeled non-$p$-packings 
cancel}
\label{sect:det-non-packings-cancel}

Partition $\myB$ into $\myB=\myP\cup\myR$, where
$\myP$ is the set of bijectively labeled $p$-packings,
and $\myR$ is the set of bijectively labeled $p$-prepackings 
that are not packings. Accordingly, 
the left-hand side of \eqref{eq:det-sieve} splits into
\[
\sum_{(\myA,\sigma,\lambda)\in\myB} m(\myA,\sigma,\lambda)
=
\sum_{(\myA,\sigma,\lambda)\in\myP} m(\myA,\sigma,\lambda)+
\sum_{(\myA,\sigma,\lambda)\in\myR} m(\myA,\sigma,\lambda)\,.
\]
We show that the rightmost sum vanishes
in characteristic 2.
To this end, it suffices to construct a fixed-point-free involution
$\phi:\myR\rightarrow\myR$ such that
$m(\phi(\myA,\sigma,\lambda))=m(\myA,\sigma,\lambda)$ holds 
for all $(\myA,\sigma,\lambda)\in\myR$.
Consider an arbitrary $(\myA,\sigma,\lambda)\in\myR$.
Since $\myA$ is a $p$-prepacking but not a packing, 
there is a minimum (with respect to e.g.~lexicographic order) 
three-tuple 
$(u_0,A_1,A_2)\in(U_3\cup U_4\cup\cdots\cup U_q)\times\myA\times\myA$ 
such that $u_0\in A_1\cap A_2$ and $A_1\neq A_2$.
Define a labeling $\lambda':d(\myA)\rightarrow L$ of $\myA$ 
by setting, for each $(u,A)\in d(\myA)$,
\begin{equation}
\label{eq:pdet-pairing}
\lambda'(u,A)=
\begin{cases}
\lambda(u,A) & \text{if $u\neq u_0$ or $A\notin\{A_1,A_2\}$;}\\
\lambda(u_0,A_2) & \text{if $u=u_0$ and $A=A_1$; and}\\
\lambda(u_0,A_1) & \text{if $u=u_0$ and $A=A_2$.}\\
\end{cases}
\end{equation}
Note that $\lambda'$ is bijective and that $\lambda'\neq\lambda$.
From \eqref{eq:pdet-monomial} and \eqref{eq:pdet-pairing}
it follows that $m(\myA,\sigma,\lambda')=m(\myA,\sigma,\lambda)$ 
holds for all $(\myA,\sigma,\lambda)\in\myR$.
We can now set $\phi(\myA,\sigma,\lambda)=(\myA,\sigma,\lambda')$
and observe that $\phi(\myA,\sigma,\lambda)\in\myR$, 
$\phi(\myA,\sigma,\lambda)\neq(\myA,\sigma,\lambda)$, and 
$\phi^2(\myA,\sigma,\lambda)=(\myA,\sigma,\lambda)$
for all $(\myA,\sigma,\lambda)\in\myR$. 
Thus, $\phi$ is a fixed-point-free involution on $\myR$.

\subsection{The algorithm}
From \eqref{eq:det-sieve} and \S\ref{sect:det-non-packings-cancel} 
we have
\begin{equation}
\label{eq:det-poly}
\sum_{(\myA,\sigma,\lambda)\in\myP} m(\myA,\sigma,\lambda)
=\sum_{J\subseteq L}(-1)^{|J|}\sum_{(\vec A,\lambda)\in\myL[J]} m(\myA,\sigma,\lambda)\,.
\end{equation}

From \eqref{eq:det-monomial} we observe that 
the left-hand side of \eqref{eq:det-poly}
is a multivariate polynomial of degree at most $pq+r$.
It follows from Lemma~\ref{lem:det-identifiability} 
that the polynomial is not identically zero if and only if 
$\myF$ contains a $p$-packing. 
It remains to evaluate \eqref{eq:det-poly} for an 
assignment of values to the indeterminates.

Let $J\subseteq L$ be fixed.
Introduce an $r\times r$ matrix $E(J)$ as follows.
Index the rows with $U_1$ and the columns with $U_2$.
Define the entry at row $u_1\in U_1$, column $u_2\in U_2$ by
\[
e_{u_1,u_2}(J)=
y_{u_1,u_2}\biggl(1+w\sum_{A\in\myF:\{u_1,u_2\}\subseteq A}x_A\prod_{u\in A\setminus\{u_1,u_2\}}\sum_{\ell\in L\setminus J}z_{u,\ell}\biggr)\,.
\]
Denote by $\myL_j[J]$ the subset of labeled
$j$-prepackings whose labeling avoids each label in $J$.
From \eqref{eq:det-formula}, \eqref{eq:det-domain}, and 
\eqref{eq:det-monomial} it immediately follows that we have
\begin{equation}
\label{eq:det-avoid}
\det E(J)=\sum_{j=0}^r\sum_{(\myA,\sigma,\lambda)\in\myL_j[J]}
 m(\myA,\sigma,\lambda)\,.
\end{equation}

Consequently, for any given assignment of values in $\F_{2^b}$
to the indeterminates $x_{A}$, $y_{u_1,u_2}$, and $z_{u,\ell}$, 
we can evaluate the left-hand side of \eqref{eq:det-avoid} 
{\em as a polynomial in the indeterminate $w$} 
via \eqref{eq:det-avoid}.
Taking the sum over all $J\subseteq L$
and extracting the coefficient of the monomial $w^p$,
we obtain an evaluation of the right-hand side of \eqref{eq:det-poly}
in total $O(2^{p(q-2)}|\myF|pq^2r^4)$ arithmetic operations in $\F_{2^b}$. 
Taking $b=\lceil\log_2 2(pq+r)\rceil$, Lemma~\ref{lem:dlsz} implies that 
we witness a $p$-packing in $\myF$ (as a nonzero
evaluation of the left-hand side of \eqref{eq:det-poly}) 
with probability at least $1/2$ in time $O^*(2^{p(q-2)})$ for
polynomial size families $\myF$.

\section{A Projection--Determinant Sieve for $p$-Packings of $q$-Sets}
\label{sec: packing}

This section establishes Theorems \ref{thm: triple packing} and
\ref{thm: q-set packing}.

\subsection{Tutte's observation}

Let us recall that 
an involution is a permutation that is identical to its inverse.
In particular, the cycle decomposition of an involution
consists of fixed points and transpositions (cycles of length 2).
It follows that the involutions on a set $I$ are 
in a one-to-one correspondence with the set partitions 
of $I$ into sets of cardinality 1 and 2.
The following lemma is essentially due to Tutte~\cite{Tutte1947}.

\begin{Lem}[Tutte's Determinant--Partition Lemma]
\label{lem:tutte-det}
Let $T$ be an $m\times m$ matrix
with entries in a multivariate polynomial ring over
a field of characteristic $2$.
Index the rows and columns of $T$ by the elements of a set $I$ 
and suppose that $T$ is symmetric so that
\[
t_{ij}=
\begin{cases}
\sum_{k}s_{\{i\},k}^2   & \text{if $i=j$;}\\
\sum_{k}s_{\{i,j\},k}   & \text{if $i\neq j$}
\end{cases}
\]
holds for all $i,j\in I$. Then,
\begin{equation}
\label{eq:tutte}
\det T\ =\!\!\!
\sum_{\substack{\iota:I\rightarrow I\\\text{$\iota$ involution}}}\,
\prod_{\substack{i\in I\\i\leq \iota(i)}}\sum_k s_{\{i,\iota(i)\},k}^2\,.
\end{equation}
\end{Lem}
\begin{Proof}
Denote the set of all permutations of $I$ by $P_I$. 
Observe that a permutation $\sigma\in P_I$ is 
not an involution 
if and only if the cycle decomposition of $\sigma$ contains 
a cycle of length at least 3.
Suppose that $\nu\in P_I$ is a permutation that is
not an involution. 
Introduce an arbitrary total order on $I$, and order 
the cycles of length at least 3 in $\nu$ based on 
the least point in $I$ moved by each such cycle. 
Denote by $\nu'$ the permutation obtained from $\nu$ 
by inverting the first cycle of length at least 3 in $\nu$.
Clearly, $\nu'\neq\nu$ and $(\nu')'=\nu$. 
Now observe that because $T$ is a symmetric matrix,
for a cyclic permutation $(i_1\ i_2\ \cdots\ i_j)$ 
and its inverse $(i_j\ i_{j-1}\ \cdots\ i_1)$,
we have
\[
t_{i_1,i_2}t_{i_2,i_3}\cdots t_{i_{j-1},i_j}t_{i_j,i_1}
=t_{i_j,i_{j-1}}t_{i_{j-1},i_{j-2}}\cdots t_{i_2,i_1}t_{i_1,i_j}
\]
It follows that
\[
\prod_{i\in I}t_{i,\nu(i)}=\prod_{i\in I}t_{i,\nu'(i)}.
\]

Partition $P_I$ into $P_I=Q_I\cup R_I$, where
$Q_I$ consists of the involutions and 
$R_I$ consists of the non-involutions.
Introduce an arbitrary total order on $R_I$.
Because the determinant of $T$ is equal
to the permanent of $T$ in characteristic 2, we have
\[
\begin{split}
\det T&=\sum_{\substack{\sigma\in P_I}}\prod_{i\in I} t_{i,\sigma(i)}\\
&=\sum_{\iota\in Q_I}\prod_{i\in I} t_{i,\iota(i)}+
\sum_{\nu\in R_I}\prod_{i\in I} t_{i,\nu(i)}\\
&=\sum_{\iota\in Q_I}
\prod_{i\in I} t_{i,\iota(i)}+
\sum_{\substack{\nu\in R_I\\
                \nu<\nu'}}
\biggl(\prod_{i\in I} t_{i,\nu(i)}+\prod_{i\in I}t_{i,\nu'(i)}\biggr)\\
&=\sum_{\iota\in Q_I}\prod_{i\in I} t_{i,\iota(i)}+
\sum_{\substack{\nu\in R_I\\
                \nu<\nu'}}
2\prod_{i\in I} t_{i,\nu(i)}\\
&=\sum_{\iota\in Q_I}\prod_{i\in I} t_{i,\iota(i)}
\end{split}
\]

Thus, splitting the product over fixed and moved points of $\iota$,
and using the symmetry of $T$, we have
\[
\begin{split}
\det T
&=\sum_{\iota\in Q_I}
  \prod_{\substack{i\in I\\i=\iota(i)}}    t_{i,\iota(i)}
  \prod_{\substack{i\in I\\i\neq\iota(i)}} t_{i,\iota(i)}\\
&=\sum_{\iota\in Q_I}
  \prod_{\substack{i\in I\\i=\iota(i)}}    \sum_{k}s_{\{i\},k}^2
  \prod_{\substack{i\in I\\i<\iota(i)}} \biggl(\sum_{k}s_{\{i,\iota(i)\},k}\biggr)^2\\
&=\sum_{\iota\in Q_I}
  \prod_{\substack{i\in I\\i=\iota(i)}}    \sum_{k}s_{\{i\},k}^2
  \prod_{\substack{i\in I\\i<\iota(i)}} 
       \biggl(\sum_{k}s_{\{i\},k}^2+2\sum_{k<k'}s_{\{i,\iota(i)\},k}s_{\{i,\iota(i)\},k'}\biggr)\\
&=\sum_{\iota\in Q_I}
  \prod_{\substack{i\in I\\i=\iota(i)}}    \sum_{k}s_{\{i\},k}^2
  \prod_{\substack{i\in I\\i<\iota(i)}}    \sum_{k}s_{\{i,\iota(i)\},k}^2\\
&=
\sum_{\iota\in Q_I}
\prod_{\substack{i\in I\\i\leq \iota(i)}}\sum_k s_{\{i,\iota(i)\},k}^2\,.
\end{split}
\]
The claim follows.
\end{Proof}

Our strategy is to leverage Tutte's observation with
random projection and sieving. In particular, we witness
a $p$-packing by randomly projecting it to a set 
$U_1\subseteq U$ where Tutte's observation forces the packing 
constraint with positive probability, which allows 
us to restrict sieving to the complementary projection
into $U_2=U\setminus U_1$.

\subsection{Admissible packings and prepackings}
\label{sect:pdet-admissible}

Let $\myF$ be a set of $q$-subsets of an $n$-element universe $U$.
Partition $U$ into two disjoint sets $U=U_1\cup U_2$ with 
$|U_1|=n_1$ and $|U_2|=n_2=n-n_1$. We say that such an
ordered partition $(U_1,U_2)$ of $U$ is an 
$(n_1,n_2)$-{\em partition}.

A subset $\myA\subseteq\myF$ is 
a $p$-{\em packing} if $|\myA|=p$ 
and the sets in $\myA$ are pairwise disjoint. 
We say that $\myA$ is {\em admissible} 
if every set $A\in\myA$ satisfies $|A\cap U_1|\leq 2$.
We say that $\myA$ is a
$(p_0,p_1,p_2)$-{\em prepacking} if
\begin{itemize}
\item[(a)]
$|\myA|=p_0+p_1+p_2$;
\item[(b)]
$|\{A\in\myA:|A\cap U_1|=j\}|=p_j$ for each $j=0,1,2$; and
\item[(c)]
the sets in $\myA$ are pairwise disjoint when projected to $U_1$.
\end{itemize}
Note that a prepacking is by definition admissible.
Also, every admissible $p$-packing is 
a $(p_0,p_1,p_2)$-prepacking for some parameters $p_0+p_1+p_2=p$.
In this case we say that the $p$-packing is a 
$(p_0,p_1,p_2)$-{\em packing}.

Let us say that a $(p_0,p_1,p_2)$-prepacking $\myA$
is {\em compatible} with an involution $\iota:U_1\rightarrow U_1$
if for every $A\in\myA$ it holds that
\begin{itemize}
\item[(a)] $\iota$ fixes the point in $A\cap U_1$ if $|A\cap U_1|=1$; and
\item[(b)] $\iota$ transposes the two points in $A\cap U_1$ if $|A\cap U_1|=2$.
\end{itemize}
Note that every prepacking is compatible with at least one involution.

\subsection{Random projection}
We analyze the probability that
a given $p$-packing projects under a random $(U_1,U_2)$
into a $(p_0,p_1,p_2)$-packing.

\begin{Lem}[Admissibility]
\label{lem:pdet-admissible}
Let $\myA$ be a $p$-packing.
For an $(n_1,n_2)$-partition $(U_1,U_2)$ of $U$ 
selected uniformly at random, we have
\begin{equation}
\label{eq:pdet-admissible-prob}
\begin{split}
\Pr&\bigl(\text{$\myA$ is a $(p_0,p_1,p_2)$-packing}\bigr)=\\
&=
\binom{p}{p_1+p_2}
\binom{p_1+p_2}{p_2}
\binom{q}{1}^{p_1}
\binom{q}{2}^{p_2}
\binom{n-pq}{n_1-p_1-2p_2}
\binom{n}{n_1}^{-1}\,.
\end{split}
\end{equation}
\end{Lem}
\begin{Proof}
Among the $p$ pairwise disjoint sets in $\myA$, there are 
$\binom{p}{p_1+p_2}$ ways to select the sets that intersect
$U_1$ in 1 or 2 points, and $\binom{p_1+p_2}{p_2}$ ways 
to select among these the $p_2$ sets that intersect in 2 points.
There are $\binom{q}{1}^{p_1}\binom{q}{2}^{p_2}$ possible 
intersection patterns with $U_1$ in these selected $p_1+p_2$ sets.
There are $\binom{n-pq}{n_1-p_1-2p_2}$ ways to select 
the remaining $n_1-p_1-2p_2$ points of $U_1$ outside the 
$pq$ points of $\myA$.
\end{Proof}

{\em Remark.} A nonzero probability is 
allocated if and only if
$pq\leq n$, $p_0+p_1+p_2=p$, $p_1+2p_2\leq n_1$, 
and $n_1-p_1-2p_2\leq n-pq$.

Using techniques similar to \S\ref{sect:projection}, 
let us derive an asymptotic approximation for \eqref{eq:pdet-admissible-prob}.
One verifies by direct calculation that for $\delta=(p_1+2p_2)/(pq)$,
\[
\bangle{n-pq}{\delta n-p_1-2p_2}\bangle{n}{\delta n}^{-1}
=\bangle{pq}{p_1+2p_2}^{-1}\,.
\]
For $n_1=\lfloor\delta n\rfloor$ thus,
uniformly for all $0<p_0,p_1,p_2<p$ with $p_0+p_1+p_2=p$,
\begin{equation}
\label{eq:pdet-adm-prob-bound}
\begin{split}
\Pr&\bigl(\text{$\myA$ is a $(p_0,p_1,p_2)$-packing}\bigr)=\\
&=
\Theta^*\biggl(
\bangle{p}{p_1+p_2}
\bangle{p_1+p_2}{p_2}
\binom{q}{1}
\binom{q}{2}
\bangle{pq}{p_1+2p_2}^{-1}
\biggr)\,.
\end{split}
\end{equation}

\subsection{Fingerprinting and identifiability}

Let  $\myA\subseteq\myF$ be a $(p_0,p_1,p_2)$-prepacking.
The {\em domain} of the prepacking is the set
\begin{equation}
\label{eq:pdet-domain}
d(\myA)
=\{(u,A):u\in A\in\myA\}
\subseteq U_2\times\myF\,.
\end{equation}
Observe that $|d(\myA)|=qp_0+(q-1)p_1+(q-2)p_2$.

Let $L$ be a set of $qp_0+(q-1)p_1+(q-2)p_2$ labels.
A {\em labeling} of $\myA$ 
is a pair $(\iota,\lambda)$, 
where $\iota:U_1\rightarrow U_1$ is an involution
compatible with $\myA$ and $\lambda:d(\myA)\rightarrow L$ 
is an arbitrary mapping. The labeling is {\em bijective}
if $\lambda$ is a bijection. 
We say that a triple $(\myA,\iota,\lambda)$ is 
a {\em labeled} $(p_0,p_1,p_2)$-prepacking.

The sieve operates over a multivariate polynomial ring with 
the coefficient field $\F_{2^b}$ and the following indeterminates. 
Introduce the indeterminates $w_0$, $w_1$, and $w_2$ 
for tracking the parameters $p_0,p_1,p_2$ of $\myA$. 
Associate with each 
$A\in\myF$ an indeterminate $x_A$.
Associate with each set $K\subseteq U_1$ 
of size $1\leq |K|\leq 2$ an indeterminate $y_K$.
Associate with each pair $(u,\ell)\in U_2\times L$ 
an indeterminate $z_{u,\ell}$.

The {\em monomial} of a labeled $(p_0,p_1,p_2)$-prepacking 
$(\myA,\iota,\lambda)$ is
\begin{equation}
\label{eq:pdet-monomial}
\begin{split}
m(\myA,\iota,\lambda)=
w_0^{2p_0}w_1^{2p_1}w_2^{2p_2}\,\prod_{A\in\myA}\, x_{A}^2
\,\prod_{\substack{i\in U_1\\i\leq \iota(i)}}\, y_{\{i,\iota(i)\}}^2
\!\!\!\prod_{(u,A)\in d(\myA)}\!\!\! z_{u,\lambda(u,A)}^2\,.
\end{split}
\end{equation}

\begin{Lem}[Identifiability]
\label{lem:pdet-identifiability}
The monomial $m(\myA,\iota,\lambda)$ uniquely determines 
both $\myA$ and $\iota$. Furthermore, 
if $\myA$ is a $p$-packing and $\lambda$ is bijective, 
then $m(\myA,\iota,\lambda)$ uniquely determines $\lambda$.
\end{Lem}

\subsection{Sieving for bijective labelings}

Denote by $\myL_{p_0,p_1,p_2}$ the set of all labeled 
$(p_0,p_1,p_2)$-prepackings.
For $J\subseteq L$, denote by $\myL_{p_0,p_1,p_2}[J]$ the subset 
of labeled $(p_0,p_1,p_2)$-prepackings whose labeling {\em avoids} 
each label in $J$. Denote by $\myB_{p_0,p_1,p_2}$ the set of all 
bijectively labeled $(p_0,p_1,p_2)$-prepackings.

By the principle of inclusion-exclusion,
\begin{equation}
\label{eq:pdet-sieve}
\sum_{(\myA,\iota,\lambda)\in\myB_{p_0,p_1,p_2}} m(\myA,\iota,\lambda)
=
\sum_{J\subseteq L}(-1)^{|J|}\sum_{(\myA,\iota,\lambda)\in\myL_{p_0,p_1,p_2}[J]}
 m(\myA,\iota,\lambda)\,.
\end{equation}

\subsection{Fingerprints of bijectively labeled non-$(p_0,p_1,p_2)$-packings 
cancel}
\label{sect:pdet-non-packings-cancel}

Let $\myB_{p_0,p_1,p_2}$ be
the set of bijectively labeled $(p_0,p_1,p_2)$-prepackings.
Partition $\myB_{p_0,p_1,p_2}$
into 
$\myB_{p_0,p_1,p_2}=\myP_{p_0,p_1,p_2}\cup\myR_{p_0,p_1,p_2}$, where
$\myP_{p_0,p_1,p_2}$ is the set of bijectively 
labeled $(p_0,p_1,p_2)$-packings,
and $\myR_{p_0,p_1,p_2}$ is the set of bijectively labeled 
$(p_0,p_1,p_2)$-prepackings that are not $(p_0,p_1,p_2)$-packings. 
Accordingly, we have
\[
\sum_{(\myA,\iota,\lambda)\in\myB_{p_0,p_1,p_2}}\!\!\!\! m(\myA,\iota,\lambda)
=
\!\!\!\!\sum_{(\myA,\iota,\lambda)\in\myP_{p_0,p_1,p_2}}\!\!\!\! m(\myA,\iota,\lambda)+
\!\!\!\!\sum_{(\myA,\iota,\lambda)\in\myR_{p_0,p_1,p_2}}\!\!\!\! m(\myA,\iota,\lambda)\,.
\]
By a pairing argument essentially identical to the one given 
in \S\ref{sect:det-non-packings-cancel}, the rightmost sum 
vanishes in characteristic 2.


\subsection{The algorithm}

Let us assume that the parameters $0<p_0,p_1,p_2<p$ and $n_1,n_2$ 
have been fixed so that any given $p$-packing is a $(p_0,p_1,p_2)$-packing
with positive probability. (We will set the precise values in what follows.)
The algorithm repeats the following randomized procedure.

First, the procedure selects an ordered $(n_1,n_2)$-partition
$(U_1,U_2)$ uniformly at random among all the $\binom{n}{n_1}$ such
partitions.

Next, the procedure evaluates the following generating function
for a random assignment of values to the indeterminates.
From \eqref{eq:pdet-sieve} and \S\ref{sect:pdet-non-packings-cancel} 
we have
\begin{equation}
\label{eq:pdet-poly}
\sum_{(\myA,\iota,\lambda)\in\myP_{p_0,p_1,p_2}}\!\!\!\! m(\myA,\iota,\lambda)
=
\sum_{J\subseteq L}(-1)^{|J|}\sum_{(\myA,\iota,\lambda)\in\myL_{p_0,p_1,p_2}[J]}
 m(\myA,\iota,\lambda)\,.
\end{equation}
The left-hand side of \eqref{eq:pdet-poly} 
is a multivariate polynomial of degree at most 
$2n_1+(2q+4)p_0+(2q+3)p_1+(2q+2)p_2$.
It follows from Lemma~\ref{lem:pdet-identifiability} 
that the polynomial is not identically zero 
if and only if $\myF$ contains a $(p_0,p_1,p_2)$-packing. 

It remains to evaluate the right-hand side of \eqref{eq:pdet-poly}.
Let $J\subseteq L$ be fixed.
The procedure relies on the following observation 
that a sum over labeled prepackings factors
into a product of two independent expressions.
The first expression, $S(J)$, generates the sets that 
do not intersect $U_1$ with a simple product.
The second expression, $\det T(J)$, generates the sets that
intersect $U_1$ with Lemma~\ref{lem:tutte-det}. 

In precise terms, let
\begin{equation}
\label{eq:pdet-s}
S(J)=\prod_{\substack{A\in\myF\\A\cap U_1=\emptyset}}\biggl(1+w_0^2x_A^2\prod_{u_2\in A}\sum_{\ell\in L\setminus J}z_{u_2,\ell}^2\biggl)\,.
\end{equation}
Define the symmetric $n_1\times n_1$ matrix $T(J)$
as follows. Index the rows and columns by elements of $U_1$.
For $u_1\in U_1$, define the diagonal entries by 
\begin{equation}
\label{eq:pdet-t-diag}
\,t_{u_1,u_1}(J)=
y_{\{u_1\}}^2
\biggl(1+w_1^2
   \!\!\!\!\sum_{\substack{A\in\myF\\A\cap U_1=\{u_1\}}}\!\!\!\!
      x_A^2\,\prod_{u_2\in A\cap U_2}\,\sum_{\ell\in L\setminus J}\,z_{u_2,\ell}^2
        \biggr)\,.
\end{equation}
For $u_1,v_1\in U_1$ with $u_1\neq v_1$, define the off-diagonal
entries by
\begin{equation}
\label{eq:pdet-t-offdiag}
\,t_{u_1,v_1}(J)=
y_{\{u_1,v_1\}}
\biggl(1+w_2
   \!\!\!\!\sum_{\substack{A\in\myF\\A\cap U_1=\{u_1,v_1\}}}\!\!\!\!
      x_A\,\prod_{u_2\in A\cap U_2}\,\sum_{\ell\in L\setminus J}\,z_{u_2,\ell}
        \biggr)\,.
\end{equation}

Let us now observe that
\begin{equation}
\label{eq:pdet-s-t}
\sum_{0\leq p_0\leq |\myF|}\ 
\sum_{0\leq p_1+2p_2\leq n_1}\ 
\sum_{(\myA,\iota,\lambda)\in\myL_{p_0,p_1,p_2}[J]}\ 
m(\myA,\iota,\lambda)=
\,S(J)\,\det T(J)\,.
\end{equation}
To this end, 
first recall \eqref{eq:pdet-monomial} and the
notion of compatibility between a prepacking 
and an involution (\S\ref{sect:pdet-admissible}). 
Next, expand \eqref{eq:pdet-t-diag} and \eqref{eq:pdet-t-offdiag} 
to sums of monomials and apply Lemma~\ref{lem:tutte-det} 
to conclude that $\det T(J)$ is exactly the left-hand side of 
\eqref{eq:pdet-s-t} restricted to $p_0=0$. 
Finally, expand \eqref{eq:pdet-s} to conclude that 
\eqref{eq:pdet-s-t} holds.

Consequently, for any given assignment of values in $\F_{2^b}$
to the indeterminates $x_{A}$, $y_K$, and $z_{u,\ell}$, 
the procedure evaluates the left-hand side of \eqref{eq:pdet-s-t} 
{\em as a polynomial in the indeterminates $w_0,w_1,w_2$} 
using a total of $O(|\myF|^6q|L|n_1^3)$ arithmetic operations 
in $\F_{2^b}$. 
From such an evaluation we can recover the coefficient
of the monomial $w_0^{p_0}w_1^{p_1}w_2^{p_2}$. This
coefficient corresponds to an evaluation of the inner sum 
in the right-hand side of \eqref{eq:pdet-poly}.
Taking the sum over $J\subseteq L$ (and multiplying 
by $w_0^{p_0}w_1^{p_1}w_2^{p_2}$), we obtain an evaluation
of the right-hand side of \eqref{eq:pdet-poly}.

Denoting the probability that a $p$-packing $\myA$ is
a $(p_0,p_1,p_2)$-packing with $P(n,n_1,p_0,p_1,p_2)$,
and taking $r=\lceil 1/P(n,n_1,p_0,p_1,p_2)\rceil$ 
repetitions of the procedure with $b=\lceil\log_2 16n\rceil$,
Lemma~\ref{lem:dlsz} implies that at least one repetition 
of the procedure witnesses any fixed $p$-packing $\myA$ 
(as a nonzero evaluation of \eqref{eq:pdet-poly}) 
with probability at least $(1-e^{-1})/2$ in time
\[
O\bigl(2^{qp_0+(q-1)p_1+(q-2)p_2}|\myF|^6pq^2n_1b^2/P(n,n_1,p_0,p_1,p_2)\bigr).
\]
Setting $n_1=\lfloor \delta n \rfloor$, $p_1=\lfloor \beta_1p\rfloor$,
$p_2=\lfloor \beta_2p\rfloor$, and $p_0=p-p_1-p_2$, we obtain from
\eqref{eq:pdet-adm-prob-bound} the running time
\[
O^*\biggl(\biggl(
\frac{2^{q(1-\beta_1-\beta_2)+(q-1)\beta_1+(q-2)\beta_2}
      \bangle{q}{\beta_1+2\beta_2}}
{\bangle{1}{\beta_1+\beta_2}
\bangle{\beta_1+\beta_2}{\beta_2}
\binom{q}{1}^{\beta_1}
\binom{q}{2}^{\beta_2}}
\biggr)^p\biggr)\,
\]
for polynomial size families $\myF$.
In particular, we obtain 
time $O^*(3.3432^p)$ for $q=3$
with $\beta_1=0.281509$ and $\beta_2=0.679622$,
time $O^*(7.2562^p)$ for $q=4$
with $\beta_1=0.323262$ and $\beta_2=0.612790$,
time $O^*(15.072^p)$ for $q=5$
with $\beta_1=0.338614$ and $\beta_2=0.582673$.

\section{A Determinant Sieve for Edge-Coloring}
\label{sec: coloring}

This section establishes Theorem \ref{thm: edge coloring}.

\subsection{Tutte's observation revisited}

The edges of $G$ can be colored 
with $d$ colors if and only if there exists 
a set of $d-1$ pairwise edge-disjoint perfect 
matchings in $G$. Indeed, because the graph is $d$-regular, 
each color class must be a perfect matching.

Let us now return to Lemma~\ref{lem:tutte-det}. 
We observe that \eqref{eq:tutte} in effect gives us a
multivariate generating function for the perfect matchings
in $G$. Our strategy is to introduce $d-1$ independent
copies of this generating function and sieve for 
edge-disjointness.

\subsection{Fingerprinting and identifiability}

Let $\vec M=(M_1,M_2,\ldots,M_p)$ be an ordered $p$-tuple 
of perfect matchings in $G$. The {\em domain} of $\vec M$
is the set
\begin{equation}
\label{eq:m-dom}
d(\vec M)=\{(e,i):e\in M_i\}\subseteq  E\times \{1,2,\ldots,p\}\,.
\end{equation}
Observe that $|d(\vec M)|=pn/2$. 
Let $L$ be a set of $pn/2$ labels. A {\em labeling} of
$\vec M$ is a mapping $\lambda:d(\vec M)\rightarrow L$.
The labeling is {\em bijective} if $\lambda$ is a bijection.

The sieve operates over a multivariate polynomial ring
with the coefficient field $\F_{2^b}$ and the following
indeterminates. Associate with each pair $(e,i)\in E\times\{1,2,\ldots,p\}$
an indeterminate $x_{e,i}$. Associate with each pair $(e,\ell)\in E\times L$
an indeterminate $y_{e,\ell}$.

The {\em monomial} of a labeled $p$-tuple $(\vec M,\lambda)$ is
\begin{equation}
\label{eq:m-monomial}
m(\vec M,\lambda)=
\prod_{(e,i)\in d(\vec M)} x_{e,i}^2y_{e,\lambda(e,i)}^2\,.
\end{equation}

\begin{Lem}[Identifiability]
\label{lem:m-identifiability}
The monomial $m(\vec M,\lambda)$ uniquely determines $\vec M$. 
Furthermore, if $\vec M$ consists of pairwise edge-disjoint
perfect matchings and $\lambda$ is bijective, 
then $m(\vec M,\lambda)$ uniquely determines $\lambda$.
\end{Lem}

\subsection{Sieving for bijective labelings}

Denote by $\myL$ the set of all labeled 
$p$-tuples of perfect matchings of $G$.
For $J\subseteq L$, denote by $\myL[J]$ the subset 
of labeled $p$-tuples of perfect matchings of $G$
whose labeling {\em avoids} each label in $J$. 
Denote by $\myB$ the set of all 
bijectively labeled $p$-tuples of perfect matchings of $G$.

By the principle of inclusion-exclusion,
\begin{equation}
\label{eq:m-sieve}
\sum_{(\vec M,\lambda)\in\myB} m(\vec M,\lambda)
=
\sum_{J\subseteq L}(-1)^{|J|}\sum_{(\vec M,\lambda)\in\myL[J]}
 m(\vec M,\lambda)\,.
\end{equation}

\subsection{Fingerprints of bijectively labeled non-disjoint $p$-tuples cancel}
\label{sect:m-non-packings-cancel}

Let $\myB$ be the set of all bijectively labeled $p$-tuples of 
perfect matchings of $G$. Partition $\myB$ into 
$\myB=\myP\cup\myR$, where
$\myP$ is the set of bijectively 
labeled $p$-tuples of perfect matchings that are pairwise 
edge-disjoint, and $\myR$ is the set of bijectively labeled 
$p$-tuples of perfect matchings for which there exists
at least one edge that occurs in at least two matchings in
the tuple. Accordingly, we have
\[
\sum_{(\vec M,\lambda)\in\myB}\!\!\!\! m(\myA,\iota,\lambda)
=
\!\!\!\!\sum_{(\vec M,\lambda)\in\myP}\!\!\!\! m(\vec M,\lambda)+
\!\!\!\!\sum_{(\vec M,\lambda)\in\myR}\!\!\!\! m(\vec M,\lambda)\,.
\]
By a pairing argument essentially identical to the one given 
in \S\ref{sect:det-non-packings-cancel}, the rightmost sum 
vanishes in characteristic 2.

\subsection{The algorithm}

First, the procedure evaluates the following generating function
for a random assignment of values to the indeterminates.
From \eqref{eq:m-sieve} and \S\ref{sect:m-non-packings-cancel} 
we have
\begin{equation}
\label{eq:m-poly}
\sum_{(\vec M,\lambda)\in\myP}\!\!\!\! m(\vec M,\lambda)
=
\sum_{J\subseteq L}(-1)^{|J|}\sum_{(\vec M,\lambda)\in\myL[J]}
 m(\vec M,\lambda)\,.
\end{equation}
The left-hand side of \eqref{eq:m-poly} 
is a multivariate polynomial of degree at most 
$2pn$.
It follows from Lemma~\ref{lem:m-identifiability} 
that the polynomial is not identically zero 
if and only if $G$ has a set of $p$ pairwise edge-disjoint
perfect matchings.

It remains to evaluate the right-hand side of \eqref{eq:m-poly}.
Let $J\subseteq L$ be fixed.
The procedure relies on Tutte's Lemma (Lemma~\ref{lem:tutte-det}).
For $i=1,2,\ldots,p$ define the symmetric $n\times n$ matrix 
$T^{(i)}(J)$ as follows. Index the rows and columns 
by the vertices $V$ of $G$.
Define the entries of $T^{(i)}(J)$ for all $u,v\in V$ by
\begin{equation}
\label{eq:m-t-entry}
t_{u,v}^{(i)}(J)=
\begin{cases}
0 & \text{if $u=v$ or $\{u,v\}\notin E$;}\\
x_{\{u,v\},i}
\sum_{\ell\in L\setminus J}y_{\{u,v\},\ell}
  & \text{if $\{u,v\}\in E$.}
\end{cases}
\end{equation}
From Lemma~\ref{lem:tutte-det} we have
\begin{equation}
\label{eq:m-t}
\sum_{(\vec M,\lambda)\in\myL[J]}\ 
m(\vec M,\lambda)=
\prod_{i=1}^p \det T^{(i)}(J)\,.
\end{equation}

Consequently, for any given assignment of values in $\F_{2^b}$
to the indeterminates $x_{e,i}$ and $y_{e,\ell}$, 
the procedure evaluates the left-hand side of \eqref{eq:m-t} 
using a total of $O(pn^3)$ arithmetic operations in $\F_{2^b}$. 
Taking the sum over $J\subseteq L$, we obtain an evaluation
of the right-hand side of \eqref{eq:m-poly}. Taking
$b=\lceil \log_2 4pn\rceil$, we witness a set of 
$p$ pairwise edge-disjoint perfect matchings in $G$ 
as a nonzero evaluation of the left-hand side of \eqref{eq:m-poly}
with probability $\Omega(1)$ in time $O^*(2^{pn/2})$ 
and space polynomial in $n$. Taking $p=d-1$, we
obtain a polynomial-space randomized algorithm for 
deciding whether a $d$-regular graph admits
a coloring of its edges with $d$ colors in 
$O^*(2^{(d-1)n/2})$ time.

\subsection{Graphs that are not regular}

Let $m=|E|$. We can modify the previous algorithm 
to run in time $O^*(2^m)$ and space polynomial in $n$
on graphs that are not regular. In particular, instead
of perfect matchings consider matchings, set $|L|=m$, 
in \eqref{eq:m-t-entry} set the diagonal entries equal to 1, 
and set $p=\Delta$, where $\Delta$ is the maximum degree
of a vertex in $G$.



\renewcommand{\thepage}{\textsc{References~p.~\arabic{page}}}
\setcounter{page}{1}


\begin{thebibliography}{10}

\bibitem{AYZ1995}
N.~Alon, R.~Yuster, and U.~Zwick, 
Color-coding, 
{\em J.\ Assoc.\ Comput.\ Mach.} 42:844--856, 1995.

\bibitem{Bjork2010a}
A.~Bj\"orklund,
Exact covers via determinants,
in \emph{Proc.\ 27th International Symposium on Theoretical Aspects of
  Computer Science, STACS 2010} (Nancy, France, March 4--6, 2010), LIPIcs
  5 Schloss Dagstuhl -- Leibniz-Zentrum f\"ur Informatik, pages
  95--106, 2010.

\bibitem{Bjork2010b} A.~Bj\"orklund, Determinant sums for undirected
  Hamiltonicity,
in \emph{Proc.\ 51st Annual IEEE Symposium on Foundations
    of Computer Science, FOCS 2010} (Las Vegas, USA, October 23--26,
  2010).

\bibitem{BHK} A.~Bj\"orklund, T.~Husfeldt, and M.~Koivisto, 
  Set partitioning via inclusion--exclusion.
\emph{SIAM J.\ Comput} 39(2):546--563, 2009.

\bibitem{Bodlaender1993}
H.~L.~Bodlaender,
On linear time minor tests with depth-first search,
\emph{J.\ Algorithm.} 14(1):1--23, 1993.

\bibitem{Chen2007} J.~Chen, S.~Lu, S.-H.~Sze, and F.~Zhang, Improved
  algorithms for path, matching, and packing problems, in
  \emph{Proc.\ 18th Annual ACM--SIAM Symposium on Discrete
    Algorithms, SODA 2007} (Philadelphia, PA, USA, 2007), pages
  298--307.

\bibitem{DowneyFellows1999}
R.~G.~Downey and M.~R.~Fellows, 
\emph{Parameterized Complexity},
Springer, 1999.

\bibitem{DFK}
R.~G.~Downey, M.~R.~Fellows, and M.~Koblitz,
Techniques for exponential parameterized reductions in vertex set
problems,
unpublished, reported in \cite[\S 8.3]{DowneyFellows1999}.

\bibitem{DeMilloLipton1978}
R.~A.~DeMillo and R.~J.~Lipton,
A probabilistic remark on algebraic program testing,
{\em Inform.\ Process\ Lett.} 7:193--195, 1978.

\bibitem{Edmonds1967}
J.~Edmonds, 
Systems of distinct representatives and linear algebra,
{\em J.\ Res.\ Nat.\ Bur.\ Standards Sect.\ B} 71B:241--245, 1967.

\bibitem{FellowsPinar} M.~R.~Fellows, P.~Heggernes, F.~A.~Rosamond,
  C.~Sloper, and J.~A.~Telle, Exact algorithms for finding $k$
  disjoint triangles in an arbitrary graph, in \emph{Proc.\ 30th
    International Workshop on Graph-Theoretic Concepts in Computer
    Science, WG 2004} (Bad Honnef, Germany, June 21--23, 2004),
  Springer LNCS 3353, pages 257--269, 2004.

\bibitem{FellowsKnauer}
M.~R.~Fellows, C.~Knauer, N.~Nishimura, P.~Ragde, F.~Rosamond, U.~Stege, D.~M.~Thilikos, S.~Whitesides,
Faster fixed-parameter tractable algorithms for matching and packing problems,
\emph{Algorithmica} 52(2):167--176, 2008

\bibitem{FernauRaible}
H.~Fernau  and D.~Raible,
A parameterized perspective on packing paths of length two,
\emph{J.\ Comb.\ Optim.} 18(4):319--341, 2009.


\bibitem{GKC} P.~Golovac, D.~Kratsch, and J.-F.~Couturier, Colorings
  with few colors: Counting, enumeration and combinatorial bounds, in
  \emph{Proc.\ 36th International Workshop on Graph-Theoretic Concepts
    in Computer Science, WG} (Zaros, Crete, Greece, June 28--30,
  2010).

\bibitem{HellKirkpatrick}
P.~Hell and D.~Kirkpatrick,
On the complexity of a generalized matching problem,
in \emph{Proc.\ 10th ACM Symposium on Theory of Computing, STOC}
(San Diego, CA, USA, May 1--3, 1978), pages 309--318, 1978.

\bibitem{Holyer}
I.~Holyer,
The NP-completeness of some edge-partition problems,
\emph{SIAM J.\ Comput.} 10(4):713--717,  1981.

\bibitem{HolyerColor}
I.~Holyer,
The NP-completeness of edge-coloring, 
\emph{SIAM J.\ Comput.} 10:718--720, 1981.


\bibitem{IPZ}
R.~Impagliazzo, R.~Paturi, F.~Zane, 
Which problems have strongly exponential complexity?, 
\emph{J.\ Comput.\ Syst.\ Sci.} 63:512--530, 2001.

\bibitem{Jia2004}
W.~Jia, C.~Zhang, and J.~Chen, 
An efficient parameterized algorithm for $m$-set packing,
\emph{J.\ Algorithm.} 50(1):106--117, 2004.


\bibitem{Koutis2005}
I.~Koutis, 
A faster parameterized algorithm for set packing,
\emph{Inform.\ Process Lett.} 94:7--9, 2005.

\bibitem{Koutis2008}
I.~Koutis,
Faster algebraic algorithms for path and packing problems, 
in \emph{Proc.\ 35th
  International Colloquium on Automata, Languages and Programming, ICALP} (Reykjavik, Iceland, July 7--11,
2008), Springer LNCS 5125, pages 575--586, 2008.

\bibitem{KoutisWilliams2009}
I.~Koutis and R.~Williams,
Limits and applications of group algebras for parameterized problems,
in \emph{Proc.\ 36th
  International Colloquium on Automata, Languages and Programming,  ICALP} (Rhodes, Greece, July 5--12, 2009),
Springer LNCS 5555, pages 653--664, 2009.

\bibitem{Kowalik2008} \L.~Kowalik, Edge colouring, in F.~Fomin \emph{et
    al.} (eds.), \emph{Open Problems: Moderately Exponential Time
    Algorithms}, Dagstuhl Seminar Proceedings 08431, 2008.

\bibitem{Kowalik2009}
\L.~Kowalik, Improved edge-coloring with three colors,
\emph{Theor.\ Comput.\ Sci.} 410(38--40):3733--3742, 2009.

\bibitem{Kneis2006}
J.~Kneis, D.~M\"olle, S.~Richter, and P.~Rossmanith, 
Divide-and-color,
in \emph{Proc.\ 32nd International Workshop on Graph-Theoretic
  Concepts in Computer Science, WG} (Bergen, Norway, June 22--24,
2006), Springer LNCS 4271, pages 58--67, 2006.

\bibitem{LevenGalil}
D.~Leven and Z.~Galil,
NP completeness of finding the chromatic index of regular graphs,
\emph{J.\ Algorithm.} 4(1):35--44, 1983.

\bibitem{Liu} Y.~Liu, S.~Lu, J.~Chen, and S.-H.~Sze, Greedy
  localization and color-coding: improved matching and packing
  algorithms, in \emph{Proc.\ 2nd International Workshop on
    Parameterized and Exact Computation, IWPEC} (Z\"urich,
  Switzerland, September 13-15, 2006), Springer LNCS 4169, pages
  84--95, 2006.

\bibitem{Mathieson2004}
L.~Mathieson, E.~Prieto, and P.~Shaw,
Packing edge disjoint triangles: a parameterized view,
in \emph{Proc.\ 1st International Workshop on Parameterized and Exact
  Computation, IWPEC} (Bergen, Norway, September 14--17, 2004)
Springer LNCS 3162, pages 127--137, 2004.

\bibitem{Monien1985}
B.~Monien, 
How to find long paths efficiently, 
{\em Annals of Discrete Mathematics} 25 (1985), 239--254.

\bibitem{PapaYanna}
C.~Papadimitriou and M.~Yannakakis, 
On limited non- determinism and the complexity of the V-C dimension, 
\emph{J.\ Comput.\ Syst.\ Sci.} 53:161--170, 1996.

\bibitem{PrietoSloper2006}
E.~Prieto and C.~Sloper,
Looking at the stars,
{\em Theor.\ Comp.\ Sc.} 351(3):437--445, 2006.

\bibitem{Robbins1955}
H.~Robbins, A remark on Stirling's formula,
{\em Amer.\ Math.\ Monthly} 62:26--29, 1955.

\bibitem{Schwartz1980}
J.~T.~Schwartz, 
Fast probabilistic algorithms for verification of polynomial identities,
{\em J.\ Assoc.\ Comput.\ Mach.} 27:701--717, 1980.

\bibitem{Tutte1947}
W.~T.~Tutte,
The factorization of linear graphs,
{\em J.\ London Math.\ Soc.} 22:107--111, 1947.

\bibitem{WangFeng} J.~Wang and Q.~Feng, An $O^*(3.523^k)$
  parameterized algorithm for 3-set packing, in \emph{Proc.\ 5th
    International Conference on Theory and Applications of Models of
    Computation, TAMC} (Xi'an, China, April 25--29, 2008), Springer
  LNCS 4978, pages 82--93, 2008.

\bibitem{Wang2008}
J.~Wang, D.~Ning, Q.~Feng, and J.~Chen,
Improved  parameterized algorithm for $P_2$-packing problem (in
Chinese),
\emph{Journal of Software} 19(11):2879--2886,  2008.

\bibitem{Williams2009}
R.~Williams,
Finding paths of length $k$ in $O^*(2^k)$, 
\emph{Inform.\ Process Lett.} 109(6):301--338, 2009.

\end{thebibliography}
\end{document}